\documentclass[11pt,twoside]{article}

\usepackage{epsfig,amsfonts,color}
\usepackage{amsmath}
\usepackage{marvosym}

\usepackage{tikz}
\usepackage{tikz-3dplot}
\usetikzlibrary{calc}

\newcommand{\norm}[1]{\left| \left| #1 \right| \right|}

\usepackage{amssymb, palatino, geometry,url}
\usepackage{algorithmic}
\usepackage{algorithm}
\usepackage{booktabs}
\newcommand{\mynote}[1]{{ \color{black}{#1}}}
\usepackage[colorlinks=true,linkcolor=blue,citecolor=blue,urlcolor=blue]{hyperref}

\usepackage{subcaption}

\newcommand{\qzero}{\begin{tikzpicture}[baseline=0.2cm]
	\draw[step=0.25cm] (0,0) grid (0.5, 0.5); 
\end{tikzpicture}}
\newcommand{\qone}{\begin{tikzpicture}[baseline=0.2cm]
	\fill (0, 0.25) rectangle ++ (0.25, 0.25); 
	\draw[step=0.25cm] (0,0) grid (0.5, 0.5); 
\end{tikzpicture}}
\newcommand{\qtwo}{\begin{tikzpicture}[baseline=0.2cm]
	\fill (0, 0.25) rectangle ++ (0.25, 0.25); 
	\fill (0, 0) rectangle ++ (0.25, 0.25); 
	\draw[step=0.25cm] (0,0) grid (0.5, 0.5); 
\end{tikzpicture}}
\newcommand{\qd}{\begin{tikzpicture}[baseline=0.2cm]
	\fill (0, 0.25) rectangle ++ (0.25, 0.25); 
	\fill (0.25, 0) rectangle ++ (0.25, 0.25); 
	\draw[step=0.25cm] (0,0) grid (0.5, 0.5); 
\end{tikzpicture}}
\newcommand{\qthree}{\begin{tikzpicture}[baseline=0.2cm]
	\fill (0, 0.25) rectangle ++ (0.25, 0.25); 
	\fill (0, 0) rectangle ++ (0.25, 0.25); 
	\fill (0.25, 0) rectangle ++ (0.25, 0.25); 
	\draw[step=0.25cm] (0,0) grid (0.5, 0.5); 
\end{tikzpicture}}
\newcommand{\qfour}{\begin{tikzpicture}[baseline=0.2cm]
	\fill (0, 0.25) rectangle ++ (0.25, 0.25); 
	\fill (0, 0) rectangle ++ (0.25, 0.25); 
	\fill (0.25, 0) rectangle ++ (0.25, 0.25);
	\fill (0.25, 0.25) rectangle ++ (0.25, 0.25); 
	\draw[step=0.25cm] (0,0) grid (0.5, 0.5);  
\end{tikzpicture}}

\newcommand{\threeDq}[9]{\begin{tikzpicture}[baseline=0.125cm]
  \draw[thick](#1,#1,0)--(0,#1,0)--(0,#1,#1)--(#1,#1,#1)--(#1,#1,0)--(#1,0,0)--(#1,0,#1)--(0,0,#1)--(0,#1,#1);
  \draw[thick](#1,#1,#1)--(#1,0,#1);
  \draw[dash pattern=on 1.2pt off 1.2pt](#1,0,0)--(0,0,0)--(0,#1,0);
  \draw[dash pattern=on 1.2pt off 1.2pt](0,0,0)--(0,0,#1);
  \pgfmathsetmacro{\one}{#2}
  \node[inner sep=1.5pt,circle,draw, fill=gray!\one] at (0, #1, #1) {}; 
  \pgfmathsetmacro{\two}{#3}
  \node[inner sep=1.5pt,circle,draw, fill=gray!\two] at (#1, #1, #1) {}; 
  \pgfmathsetmacro{\three}{#4}
  \node[inner sep=1.5pt,circle,draw, fill=gray!\three] at (0, #1, 0) {}; 
  \pgfmathsetmacro{\four}{#5}
  \node[inner sep=1.5pt,circle,draw, fill=gray!\four] at (#1, #1, 0) {}; 
  \pgfmathsetmacro{\five}{#6}
  \node[inner sep=1.5pt,circle,draw, fill=gray!\five] at (0, 0, #1) {}; 
  \pgfmathsetmacro{\six}{#7}
  \node[inner sep=1.5pt,circle,draw, fill=gray!\six] at (#1, 0, #1) {}; 
  \pgfmathsetmacro{\seven}{#8}
  \node[inner sep=1.5pt,circle,draw, fill=gray!\seven] at (0, 0, 0) {}; 
  \pgfmathsetmacro{\eight}{#9}
  \node[inner sep=1.5pt,circle,draw, fill=gray!\eight] at (#1, 0, 0) {}; 
\end{tikzpicture}}

\usepackage{fancyhdr}
\usepackage{lineno}

\begin{document}

\title{A Fast and Scalable Computational Topology Framework\\  for the Euler Characteristic}

\author{Daniel Laky and Victor M. Zavala\thanks{Corresponding Author: victor.zavala@wisc.edu}\\
 {\small Department of Chemical and Biological Engineering}\\
 {\small \;University of Wisconsin-Madison, 1415 Engineering Dr, Madison, WI 53706, USA}}
 \date{}
\maketitle

\begin{abstract}
The Euler characteristic (EC) is a powerful topological descriptor that can be used to quantify the shape of data objects that are represented as fields/manifolds. Fast methods for computing the EC are required to enable processing of high-throughput data and real-time implementations. This represents a challenge when processing high-resolution 2D field data (e.g., images) and 3D field data (e.g., video, hyperspectral images, and space-time data obtained from fluid dynamics and molecular simulations).  In this work, we present parallel algorithms (and software implementations) to enable fast computations of the EC for 2D and 3D fields using vertex contributions.  We test the proposed algorithms using synthetic data objects and data objects arising in real applications such as microscopy, 3D molecular dynamics simulations, and hyperspectral images. Results show that the proposed implementation can compute the EC a couple of orders of magnitude faster than {\tt GUDHI} (an off-the-shelf and state-of-the art tool) and at speeds comparable to {\tt CHUNKYEuler} (a tool tailored to scalable computation of the EC). The vertex contributions approach is flexible in that it compute the EC as well as other topological descriptors such as perimeter, area, and volume ({\tt CHUNKYEuler} can only compute the EC). Scalability with respect to memory use is also addressed by providing low-memory versions of the algorithms; this enables  processing of data objects beyond the size of dynamic memory. All data and software needed for reproducing the results are shared as open-source code.
\end{abstract}


\section{Introduction}

Data objects that appear in the form of fields (e.g., images, video, space-time data) are common in science and engineering. These data objects can be processed using powerful algorithms such as convolution operations, Fourier transforms, and singular value decomposition to extract information and to enable reduction and visualization. Convolutional neural networks (CNNs), in particular, are a highly flexible tool that can be used to extract diverse types of feature information from field data. However, CNNs have significant scalability limitations (e.g., require repetitive convolutions to learn adequate operators) and might require large amounts of data to be trained. Moreover, attributing meaning to features extracted from CNNs is not straight-forward \cite{Jiang_LC_TDA}.

Topological data analysis (TDA) has recently emerged as a powerful framework to quantify the shape of field data \cite{Smith_TDA, Smith_EC, Jiang_LC_TDA}. A simple topological descriptor known as the Euler Characteristic (EC), in particular, has gained significant attention in diverse applications \cite{Euler}. The EC is a descriptor that captures basic topological features of binary fields (e.g., connected components, voids, holes) and can be extended to continuous fields by using filtration/percolation procedures (i.e., the EC is computed at different filtration values). The EC has also been used for analyzing data in neuroscience \cite{Kilner_EC, Chung_EC}, medical imaging \cite{Marsh_EC, Hakim_EC}, cosmology \cite{Schmalzing_EC, Pranav_EC}, and plant biology \cite{Amezquita_EC}. The EC has also been recently used as a descriptor/feature to train simple machine learning models (e.g., linear regression) that have comparable prediction accuracy to those of CNNs but that are significantly less computationally expensive to build \cite{Smith_EC, Jiang_LC_TDA,Smith_MD_EC}. 

Enabling fast computations of topological descriptors is necessary to handle field data at high resolutions, high-throughput data, and  to enable real-time applications (e.g., control). Methods for fast processing of small-scale images was proposed by Snidaro and Foresti \cite{Snidaro_EC}; specifically, they proposed to compute only the change in the EC over the filtration values. This idea was further explored in \cite{Heiss_EC, Wang_EC_GPU, Richardson_EC, Rieser_EC}. Heiss and Wagner presented an algorithm that computes the EC for 3D fields that are too large to fit into memory and provide a software implementation called {\tt CHUNKYEuler}\cite{Heiss_EC}. A parallel implementation of this method using GPUs has also been recently developed \cite{Wang_EC_GPU}. 

In this work, we provide parallel implementations of the vertex contributions methods of Snidaro and Foresti \cite{Snidaro_EC}. We highlight that a major contribution of this work is the generalization of of this method to 3D fields (including handling of non-binary fields and parallel implementation); these capabilities allow us to process a broad range of data sets arising in applications. The vertex contributions method is scalable and flexible, in that contributions can be used to compute the EC and other relevant topological descriptors such as perimeter, area, volume (the approach implemented in {\tt CHUNKYEuler} can only compute the EC).  We provide background information on the computation of the EC with an in-depth look at 2D/3D field data and level set filtration and outline the vertex contribution method.  We highlight aspects that make the proposed method highly parallelizable and describe our software implementation. In addition, we benchmark our implementations against state-of-the-art computational topology tools using synthetic data sets and data sets arising in real applications. Specifically, we analyze synthetic random fields that are systematically generated to obtain field data at different resolutions and we study data sets arising in real applications such as microscopy, molecular simulations, and hyperspectral images. Our results demonstrate that our implementation can compute the EC a couple of orders of magnitude faster than the off-the-shelf computational topology tool {\tt GUDHI} \cite{GUDHI}. We also compare times of the proposed methods to those implemented in the {\tt CHUNKYEuler} software \cite{Heiss_EC} and their GPU implementation \cite{Wang_EC_GPU}.

\section{Methodology}

\begin{figure}[!htp]
\begin{center}
\includegraphics[width=4in]{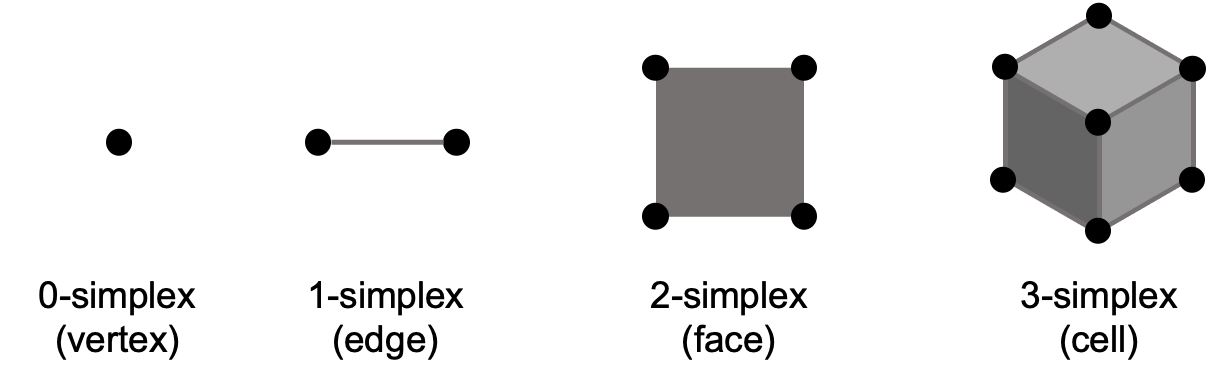}\caption{\mynote{Cubical simplexes relevant to 2D and 3D field processing.}}\label{fig:cubical_simplexes}
\end{center}
\end{figure}

\begin{figure}[!htp]
\begin{center}
\includegraphics[width=5in]{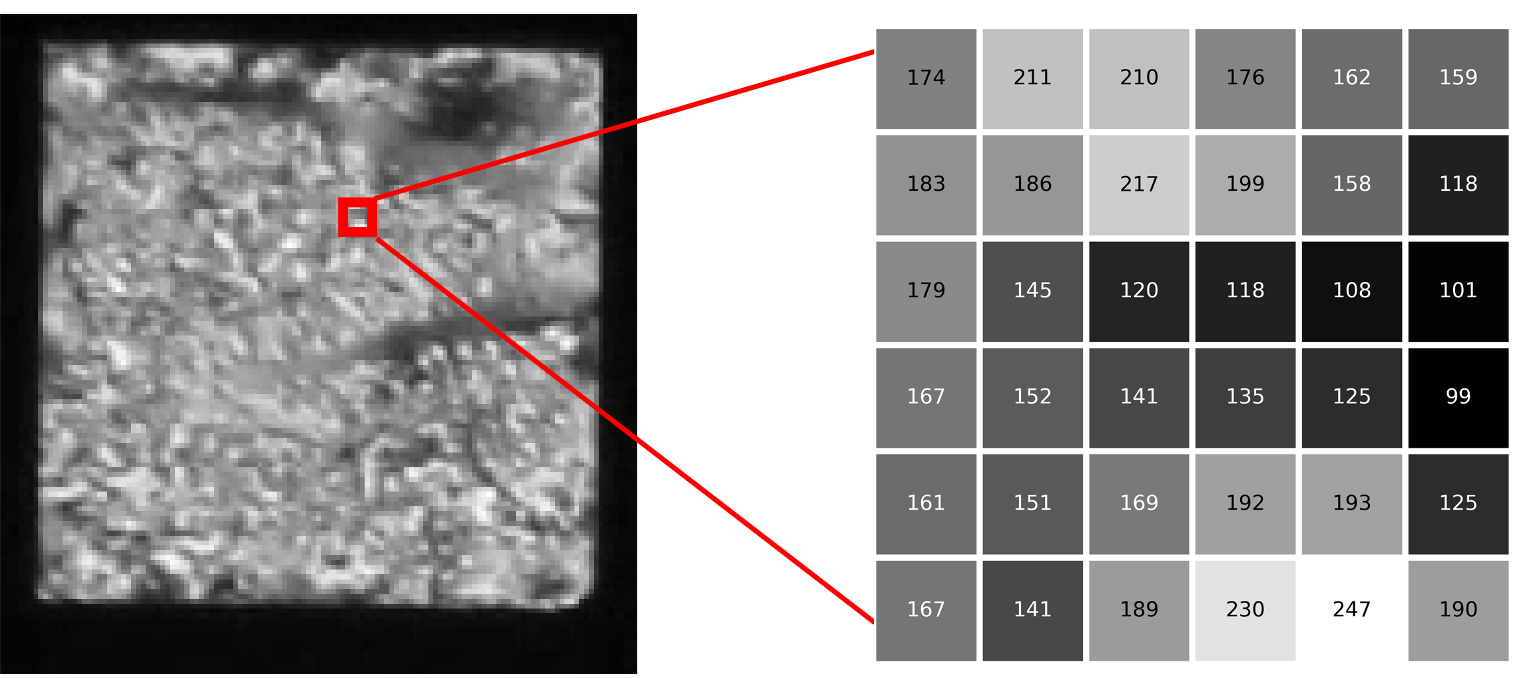}\caption{\mynote{Liquid crystal micrograph (image) represented as a cubical simplicial complex; this is done by assigning pixel values to faces which are connected by lower dimensional cubical simplexes, edges, and vertices.}}\label{fig:LC_zoom_in}
\end{center}
\end{figure}

The EC of a data object is computed by counting the contribution of each fundamental component to the overall topology of the object. These fundamental components, which we will refer to as simplexes, are the building blocks that contribute to the topology of the object. The simplexes that are important to this work, and more broadly 2D/3D field analysis, include 0-dimensional through 3-dimensional cubical simplexes (Figure \ref{fig:cubical_simplexes}). In this work we will refer to a 0-dimensional cubical simplex as a vertex, 1-dimensional as an edge, 2-dimensional as a face or pixel, and 3-dimensional as a cell or voxel. A 2D field data object is a collection of face/pixel data (e.g., images) that contains intensity information to describe what color is displayed at a particular location. By representing each pixel as a face, an image object may be represented as a collection of cubical simplexes (Figure \ref{fig:LC_zoom_in}).

The EC can be defined as the alternating sum of cubical simplexes; for fields, this is: 
\begin{align}
	\chi = V - E + F - C. \label{eqn:EC_formula}
\end{align}
Here, $V$ is the number of vertices, $E$ is the number of edges, $F$ is the number of faces/pixels, $C$ is the number of cells or voxels, and $\chi$ represents the EC value.

\begin{figure}[!htp]
\begin{center}
\includegraphics[width=5in]{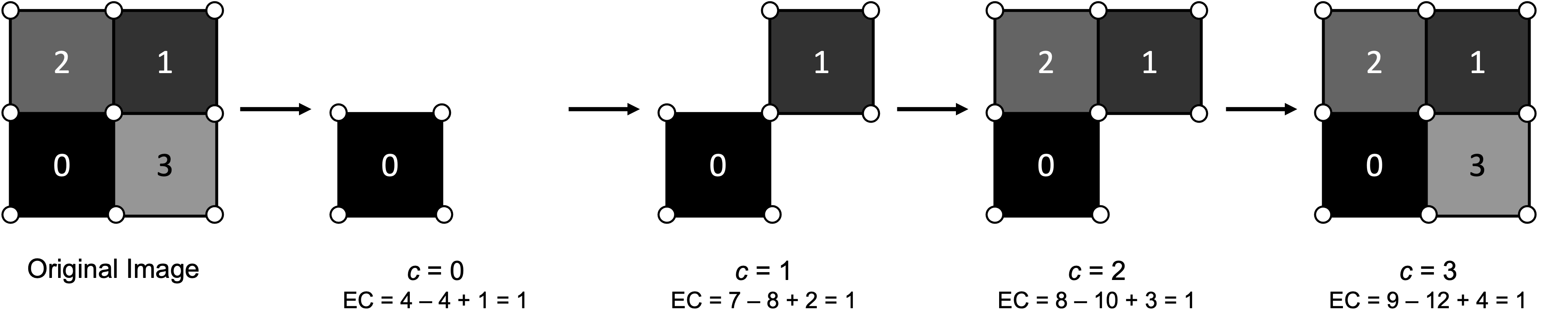}\caption{\mynote{Example 2D field undergoing the process of filtration over the full range of pixel intensity values.}}\label{fig:filtration_example}
\end{center}
\end{figure}

To generate the EC for a field (a continuous object), one must first transform the original field into a binary field by applying a filtration at a desired face/pixel intensity level. A filtration/percolation is a function that is used to define which components (e.g., pixels in the case of a 2D image) should be included and which should not. The filtration function used throughout this work is a sublevel filtration:
\begin{align}
	g^\text{-}_c\left(f\right) = \left\{\left(x_w, x_h\right) \rvert f\left(x_w, x_h\right) \leq c \right\}. \label{eqn:sublevel_set}
\end{align}
Here, $c$ is the filtration level, faces are defined at position $\left(x_w, x_h\right)$ with intensity $f$, and set $g^\text{-}_c$ contains faces that are less than or equal to $c$ only. Importantly, when a face is included, all edges and vertices relevant to that face are also included in the set. An example of filtration for multiple values of $c$ over a small image is shown in Figure \ref{fig:filtration_example}. The collection of EC values obtained at different levels $c$ is known as the EC curve. 

\begin{figure}[!htp]
\begin{center}
\includegraphics[width=5in]{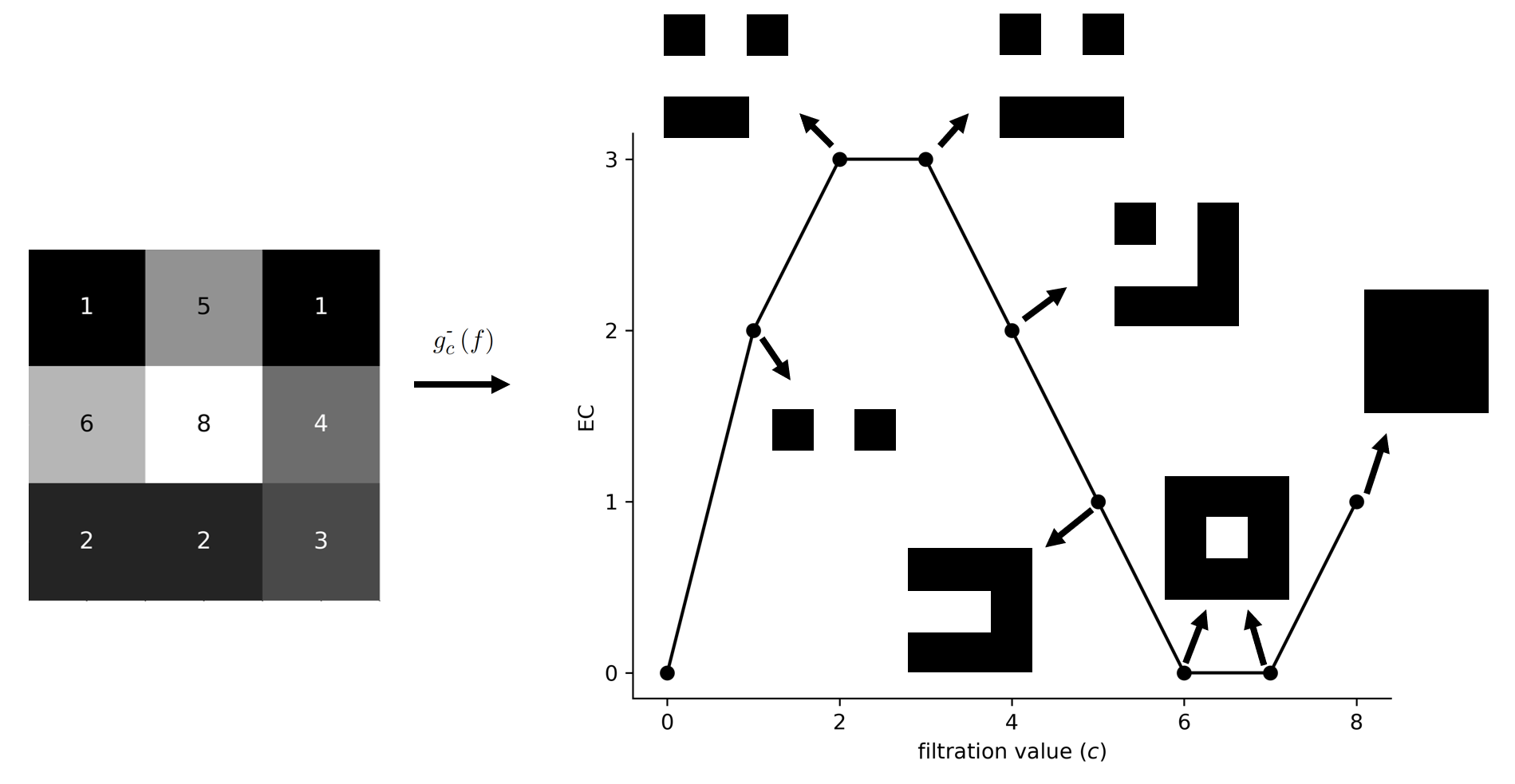}\caption{\mynote{EC curve for a 2D field. Note that, in this example, we see the emergence of a hole in the simplicial complex at a filtration level of $c = 6$.}}\label{fig:filtration_example_EC_curve}
\end{center}
\end{figure}

The filtrations in Figure \ref{fig:filtration_example} can be verified by counting the number of components included and utilizing \eqref{eqn:EC_formula}. A generalization of the EC, called the EC curve, encodes information for the entire field by applying $g^\text{-}_c$ for values of $c$ that cover the entire range of face intensities for a given field. A small example of an EC curve is shown in Figure \ref{fig:filtration_example_EC_curve} for a 2D field.

\begin{figure}[!htp]
\begin{center}
\includegraphics[width=3in]{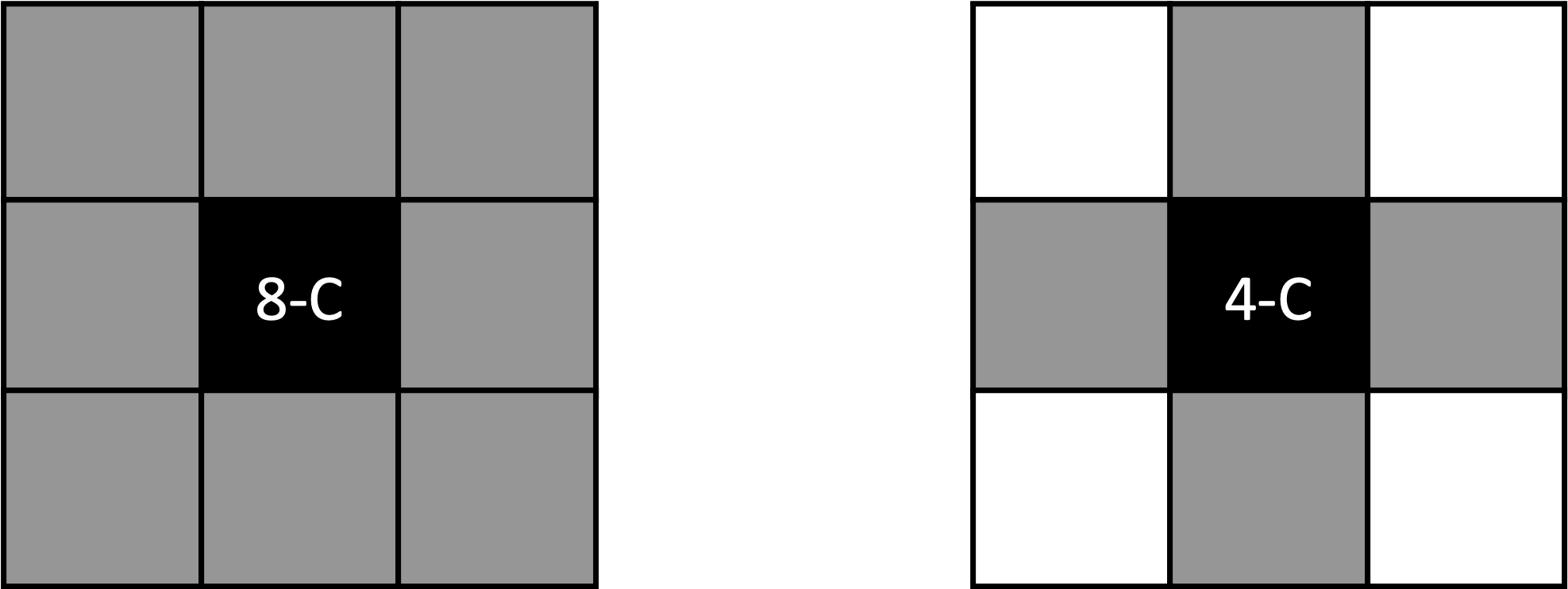}\caption{\mynote{Types of connectivity; on the left is vertex-adjacency (or 8-C) and, on the right, is edge-adjacency or 4-C.}}\label{fig:adjacency_2D}
\end{center}
\end{figure}

A fundamental aspect to consider in EC computations is connectivity. There are a couple of types of adjacency in a 2D field: (i) vertex adjacency and (ii) edge adjacency. Vertex adjacency defines that a face that shares a vertex with another face is adjacent. This is often referred to as 8-connectedness (or 8-C) as all 8 of the faces that surround an arbitrary central face are considered connected to the central face, as seen in Figure \ref{fig:adjacency_2D}. Edge adjacency is more strict, as a face is adjacent to another face only if an edge is shared. This is often referred to as 4-connectedness (or 4-C) as there are only 4 edge-connected faces from an arbitrary central face, as seen in Figure \ref{fig:adjacency_2D}. In the above example, where the EC was computed for Figure \ref{fig:filtration_example}, 8-C was assumed. Throughout this work, we will be using vertex adjacency for all dimensions: 8-C for 2D field analysis and 26-C for 3D field analysis.

\begin{figure}[!htp]
\begin{center}
\includegraphics[width=3in]{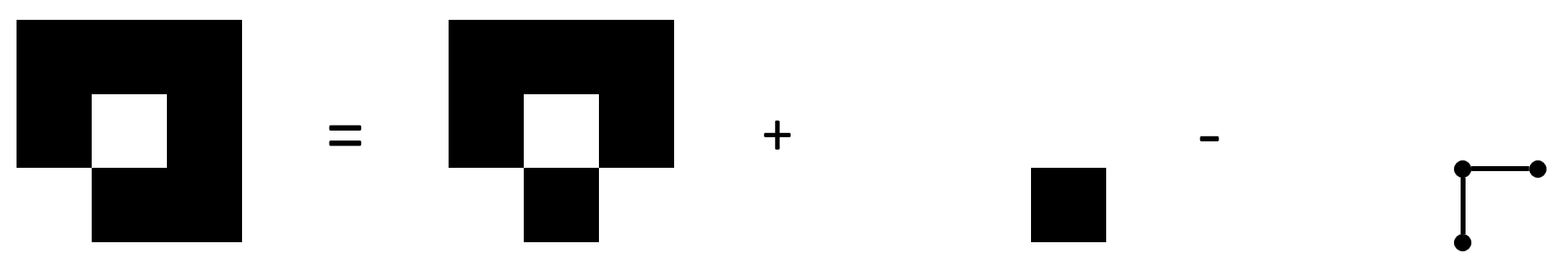}\caption{\mynote{Adding a single face to the bottom right of a 3x3 field complex requires the subtraction of two edges and 3 vertices which are double counted.}}\label{fig:inclusion_exclusion}
\end{center}
\end{figure}

A na{\"i}ve approach for computing the EC curve would be to count the vertices, edges, and faces at each filtration level. This would require iterations over every vertex, edge, and face at each level, resulting in poor computational scalability: $\mathcal{O}\left(whn_c\right)$ where $w$ and $h$ are the width and height of the field, respectively, and $n_c$ is the number of filtration levels. Another method would be to consider the vertices, edges, and faces from the previous filtration level and only add the new ones to the current filtration; unfortunately, adding a new face does not necessarily add all vertices and edges associated with the face as new components to the level set. This is because the EC follows the inclusion-exclusion principle:
\begin{align}
	\chi\left(A \cup B\right) = \chi\left(A\right) +  \chi\left(B\right) -  \chi\left(A \cap B \right). \label{eqn:inclusion_exclusion}
\end{align}
As such, the unique components added to set $A$ by adding set $B$, or a new face, would be those in $B$ minus the intersection: $A \cap B$. An illustration of this concept is included in Figure \ref{fig:inclusion_exclusion} where edges and vertices will be double-counted if we do not subtract the intersection of the new face and the existing set.

\begin{figure}[!htp]
\begin{center}
\includegraphics[width=3in]{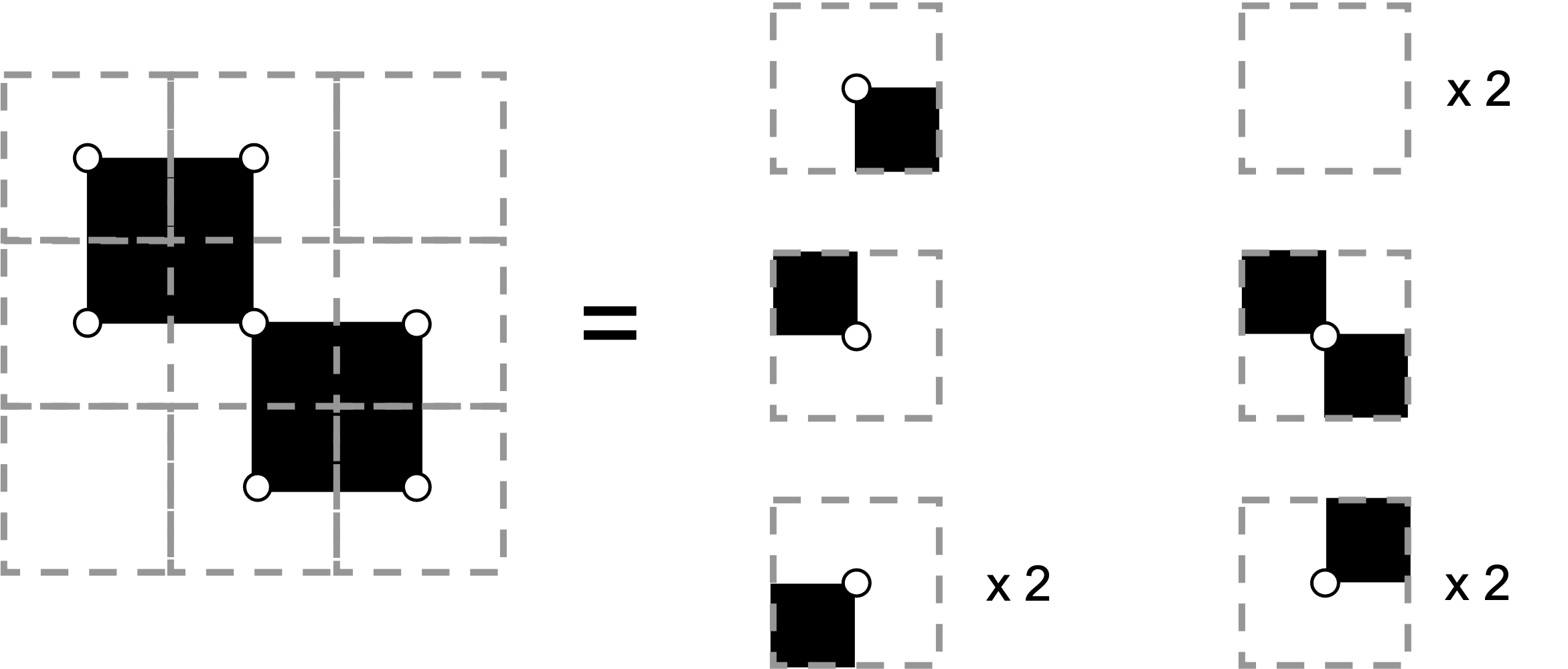}\caption{\mynote{Examining the a face-sized neighborhood about each vertex in a 2x2 binary image.}}\label{fig:vertex_contribution}
\end{center}
\end{figure}

One can avoid the aforementioned problem by tracking vertices and edges which are included in the set. This provide a means to check new vertices and edges against the existing set and only add novel components. However, this comes at large memory cost, as these additional edge and vertex arrays would require storage of a binary or integer value for each entry, exceeding the element size of the original face data array from the field data itself. A solution to both of these problems is to use bit neighborhoods to determine impact on EC \cite{Gray_bitmap}. Vertices can be used to compute EC contribution by considering a face-sized neighborhood centered on a vertex. A quarter of each face that share the vertex contribute to make up this neighborhood. Subsequently, one-half of each edge that shares the vertex are also included in this vertex-centered neighborhood. An illustration of this concept is shown in Figure \ref{fig:vertex_contribution}.

There are 16 positionally-different types of vertex contributions of which 6 are unique subject to symmetry, as shown in Table \ref{tab:2D_bitmap_contributions}. From these representations we can compute the EC contribution and we can compute the contribution to the perimeter and area of the filtered field. One advantage to this method over strictly keeping track of the EC contribution is that the perimeter cannot be computed from the edges in the set, it can only be computed from exterior edges, which would require the tracking of additional information. The contribution of these 6 types of vertex contributions is given for EC (both 4-C and 8-C), area, and perimeter in Table \ref{tab:2D_bitmap_contributions}.

\begin{table}[!htp]
\renewcommand*{\arraystretch}{1.5}
\footnotesize
\caption{The 2D vertex contributions to the overall EC for both 4-C and 8-C cases. Area and perimeter contributions (equivalent in both 4-C and 8-C cases) are given as well.}
\begin{center}
\aboverulesep = 0mm \belowrulesep = 0mm
\begin{tabular}{ccccccccccc}
\toprule
	Type 	& V (8-C) 	& E (8-C)	& F (8-C) 	& EC (8-C) & V (4-C) & E (4-C)  & F (4-C) 	& EC (4-C) & Perimeter & Area	\\ 
\midrule
	\qzero 	&     0  	&      0 	&     0 	&        0 	  &     0 	 &      0 	 &      0 		&       0   	  &       0        &   0    \\
	\qone  	&     1 	& 2 * (0.5) & 1 * (0.25) &  0.25 	  &     1 	 & 2 * (0.5) & 1 * (0.25) 	&   0.25     &       1        & 0.25 \\
	\qtwo  	&     1 	& 3 * (0.5) &2 * (0.25) &      0 	  &     1 	 & 3 * (0.5) & 2 * (0.25) 	&      0        &       1        &  0.5  \\
	\qd	  	&     1 	& 4 * (0.5) &2 * (0.25) &   -0.5 	  &     2 	 & 4 * (0.5) & 2 * (0.25) 	&     0.5      &       2        &  0.5  \\
	\qthree  	&     1 	& 4 * (0.5) &3 * (0.25) &   -0.25 	  &     1 	 & 4 * (0.5) & 3 * (0.25) 	&   -0.25     &       1        & 0.75 \\
	\qfour  	&     1 	& 4 * (0.5) &4 * (0.25) &      0 	  &     1 	 & 4 * (0.5) & 4 * (0.25) 	&      0	  &       0        &   1    \\ [1.3ex]
\bottomrule
\end{tabular}
\aboverulesep = 0.605mm \belowrulesep = 0.984mm
\end{center}
\label{tab:2D_bitmap_contributions}
\end{table}

Because there is no intersection between the contributions of each vertex, the EC can be computed by the strict addition of contributions from each vertex. In other words, \eqref{eqn:inclusion_exclusion} loses the $\chi\left(A \cap B \right)$ term because it is zero. By providing a collar of values about the edge of the field, or by determining if a given vertex is a corner, edge, or central vertex, the EC of that field may be computed by iterating over all vertices for each value of the level set. Although this is clearly an improvement over checking each vertex, edge, and face, this method still is computationally expensive in large-scale fields as the order of computation remains at $\mathcal{O}\left(whn_c\right)$. However, by tracking only the change in contribution types (i.e., the 6 unique vertex contribution types) for a given level set, time complexity can be reduced from $\mathcal{O}\left(whn_c\right)$ to $\mathcal{O}\left(wh + n_c\right)$, as shown in Snidaro and Foresti \cite{Snidaro_EC}. 

The method implemented in {\tt CHUNKYEuler} yields the same reduction in time complexity by analyzing change in EC through changes in which top-dimensional cells are active, however, only the changes in EC are tracked. Modifications would need to be made to {\tt CHUNKYEuler} to consider other topological descriptors  proposing new challenges when considering indirect metrics such as perimeter. {\tt GUDHI} computes the so-called persistence homology of a cubical complex over filtration values using the compressed annotation matrix method \cite{Boissonnat_PD} and currently cannot benefit from the reduction in time complexity by looking at changes instead of evaluating each filtration level. The limitations of {\tt GUDHI} (in terms of computational speed) and {\tt CHUNKYEuler} (in terms of ability to compute diverse descriptors) are overcome by vertex contribution algorithms.

\begin{figure}[!htp]
\begin{center}
\includegraphics[width=4.5in]{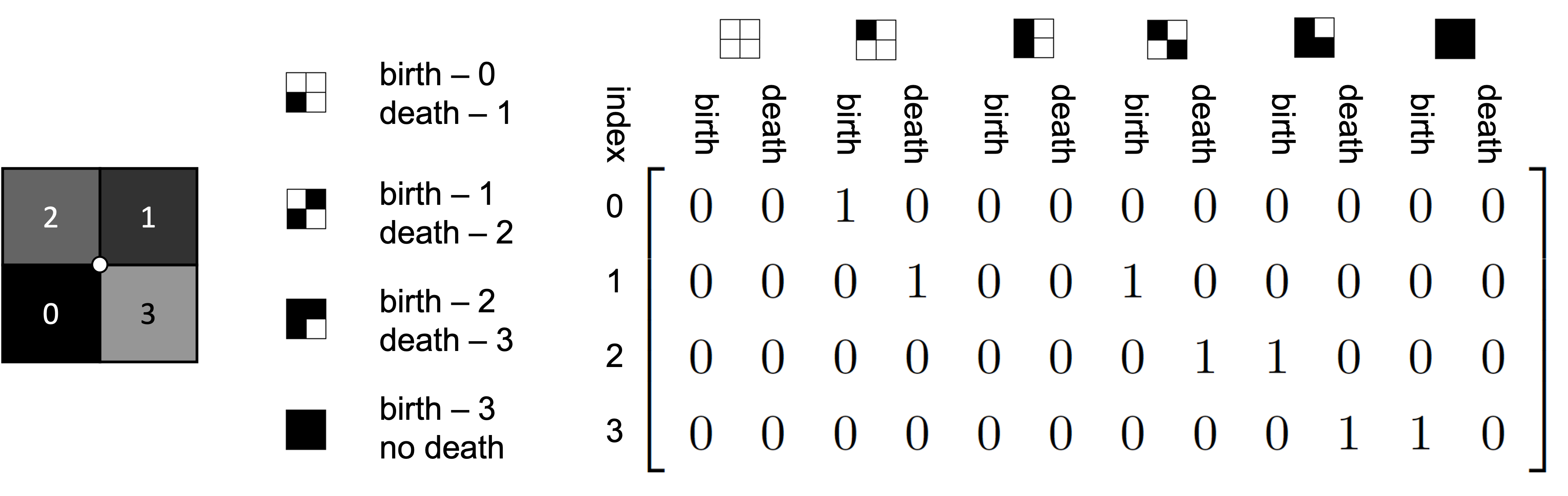}\caption{\mynote{Visualization of the contribution of a single vertex over its face values.}}\label{fig:bitmap_contribution_example}
\end{center}
\end{figure}

With our implementation of the vertex contribution method, an array that includes two entries for each contribution type at each level set value is initialized at zero. Then, for each vertex neighborhood, the integer values of the pixels are used as the index to be incremented. As an example, in Figure \ref{fig:bitmap_contribution_example} we show the  central face neighborhood of Figure \ref{fig:filtration_example}; the first face becomes active at a value of 0. Subsequently, the vertex contribution with one active face is incremented as born at index 0. Since the second face becomes active at a value of 1, the one-face vertex contribution is incremented as dying at index 1, and the two-face diagonal vertex contribution is incremented as being born. This process continues for each face value until the maximum value is reached. At this point, the neighborhood will be filled under the assumption that the maximum value for filtration exceeds the maximum intensity of faces in the field. The only additional step is determining if the first two face values are in a diagonal position or not as the two types of 2-face neighborhoods have different contributions to the topological descriptors. In the example in Figure \ref{fig:bitmap_contribution_example}, it can be seen that there will be a diagonal vertex neighborhood as the first two active faces are diagonally adjacent. Importantly, the output of this method is not just the EC curve, but rather an array of the quantity of vertex contributions of each type that are present at each level set. This means that one can compute  EC, perimeter, area, and any other descriptors/metrics that can utilize these vertex contributions as input for each filtration value. The process of gathering vertex contributions of a 2D field considering filtration values using this method is summarized in Algorithm \ref{alg:2D_contributions_serial}. In the following sections, we propose algorithms for computing vertex contributions and extend the method to 3D fields. 

\begin{algorithm}
\caption{Bitmap contribution (2D, serial)} \label{alg:2D_contributions_serial}
\begin{algorithmic}[1]
	\STATE Field $\gets$ read(raw data file)
	\STATE$w \gets$ field width
	\STATE$h \gets$ field height
	\STATE$M \gets$ largest filtration value
	\STATE$q_{k,\text{in}}, q_{k,\text{out}} \gets$ Zeros of length M+2 $\forall k \in \{0, 1, ..., 9\}$
	\FORALL{$i < h + 1$}
		\FORALL{$j < w + 1$}
			\STATE data $\gets [M+1, M+1, M+1, M+1]$ 
			\IF{$i > 0 \land j > 0$} \label{line:2D_data_start}
				\STATE data$[0] \gets$ Field$(i - 1, j - 1)$
			\ELSIF{$i > 0 \land j < w$}
				\STATE data$[1] \gets$ Field$(i - 1, j)$
			\ENDIF
			\IF{$i < h \land j > 0$}
				\STATE data$[2] \gets$ Field$(i, j - 1)$
			\ELSIF{$i < h \land j < w$}
				\STATE data$[3] \gets$ Field$(i, j)$
			\ENDIF \label{line:2D_data_stop}
			\STATE pos $\gets$ argsort(data)
			\STATE adjacency $\gets$ distance(pos$[0]$, pos$[1]$)
			\STATE $q_{1,\text{in}}[\text{data}[0]]\texttt{++}, q_{1,\text{out}}[\text{data}[1]]\texttt{++}$
			\IF{adjacency \texttt{==} 1}
				\STATE $q_{2,\text{in}}[\text{data}[1]]\texttt{++}, q_{2,\text{out}}[\text{data}[2]]\texttt{++}$
			\ELSE
				\STATE $q_{d,\text{in}}[\text{data}[1]]\texttt{++}, q_{d,\text{out}}[\text{data}[2]]\texttt{++}$
			\ENDIF
			\STATE $q_{3,\text{in}}[\text{data}[2]]\texttt{++}, q_{3,\text{out}}[\text{data}[3]]\texttt{++}$
			\STATE $q_{4,\text{in}}[\text{data}[3]]\texttt{++}$
		\ENDFOR
	\ENDFOR
\end{algorithmic}
\end{algorithm}

\subsection{Parallel Implementation}

Parallel implementation follows the same pattern as the serial implementation but requires: either (i) a vertex
contribution array for each thread, or (ii) a lock-based structure for a global vertex contribution array. The length of the 2D vertex contribution array would be one more than the number of levels, and the width would be 10. Although there are 6 vertex contribution types which would result in an array width of 12, the empty type contributes nothing to perimeter, area, and EC (see Table \ref{tab:2D_bitmap_contributions}); in addition, the number of empty contributions at a given level can be back-calculated using the total number of vertices. In addition, when considering lock-based methods, at the entry value, every vertex will increment the empty contribution at the initial filtration level, causing a pile-up that could exceed serial computation time. For instance, a standard 8-bit image with pixel values ranging from 0-255 would require 2570 entries in a vertex contribution matrix, requiring 10280 bytes of storage per array instance using 32-bit integers. In most CPU systems, holding these arrays in memory is not expensive and can be done easily for most modern systems, even with one storage array per core for highly distributed systems. The size of the image in memory will greatly exceed that of vertex contributions in most cases. For perspective, if we consider a standard high-definition (HD) image size of 1280x720, the storage for only one channel using arrays of unsigned 8-bit integers would require 921600 bytes of memory, which exceeds the size of approximately 90 vertex contribution arrays. If a vertex contribution array can be afforded for each thread, Algorithm \ref{alg:2D_contributions_parallel} can and should be used. The lock-based method for this system will not be described in detail in this work, however in highly distributed systems, such as GPUs, a global contribution array with locks should be used where local or register memory does not exceed the size of a thread vertex contribution array plus dynamic memory required to compute vertex contributions. Instead of increasing the global array values directly, the algorithm would use a locking system for each increment  operation to prevent a data race. A light-weight locking mechanism would be best for these purposes, such as a spin-lock, which would reduce overhead of waking up a thread as in a mutex-based locking system.

\begin{algorithm}
\caption{Bitmap contribution (2D, parallel)} \label{alg:2D_contributions_parallel}
\begin{algorithmic}[1]
	\STATE Field $\gets$ read(Field file)
	\STATE$M \gets$ largest filtration value
	\STATE$n_\text{threads} \gets$ number of threads
	\STATE$q_{tk,\text{in}}, q_{tk,\text{out}} \gets$ Zeros of length M+2 $\forall k \in \{0, 1, ..., 9\}, \forall t \in \{1, 2, ..., n_\text{threads}\}$
	\STATE jobs$_t \gets$ total jobs / $n_\text{threads}$
	\FORALL{$t < n_\text{threads}$}
		\FORALL{$(i, j) \in \text{jobs}_t$}
			\STATE data $\gets [M+1, M+1, M+1, M+1]$ 
			\IF{$i > 0 \land j > 0$} \label{line:2D_data_start_parallel}
				\STATE data$[0] \gets$ Field$(i - 1, j - 1)$
			\ELSIF{$i > 0 \land j < w$}
				\STATE data$[1] \gets$ Field$(i - 1, j)$
			\ENDIF
			\IF{$i < h \land j > 0$}
				\STATE data$[2] \gets$ Field$(i, j - 1)$
			\ELSIF{$i < h \land j < w$}
				\STATE data$[3] \gets$ Field$(i, j)$
			\ENDIF \label{line:2D_data_stop_parallel}
			\hspace{17em}\smash{$\left.\rule{0pt}{11.25\baselineskip}\right\}\ \mbox{in parallel}$}
			\STATE pos $\gets$ argsort(data)
			\STATE adjacency $\gets$ distance(pos$[0]$, pos$[1]$)
			\STATE $q_{t1,\text{in}}[\text{data}[0]]\texttt{++}, q_{t1,\text{out}}[\text{data}[1]]\texttt{++}$
			\IF{adjacency \texttt{==} 1}
				\STATE $q_{t2,\text{in}}[\text{data}[1]]\texttt{++}, q_{t2,\text{out}}[\text{data}[2]]\texttt{++}$
			\ELSE
				\STATE $q_{td,\text{in}}[\text{data}[1]]\texttt{++}, q_{td,\text{out}}[\text{data}[2]]\texttt{++}$
			\ENDIF
			\STATE $q_{t3,\text{in}}[\text{data}[2]]\texttt{++}, q_{t3,\text{out}}[\text{data}[3]]\texttt{++}$
			\STATE $q_{t4,\text{in}}[\text{data}[3]]\texttt{++}$
		\ENDFOR
	\ENDFOR
	\STATE join threads
	\STATE $q_{k,\text{in}} \gets \sum_{t=0}^{n_\text{threads}} q_{tk,\text{in}} \forall k \in \{0, 1, ..., 9\}$
	\STATE $q_{k,\text{out}} \gets \sum_{t=0}^{n_\text{threads}} q_{tk,\text{out}} \forall k \in \{0, 1, ..., 9\}$
\end{algorithmic}
\end{algorithm}

\subsection{3D Field Processing}

For processing 3D fields we need to define new types of contributions. To the best of our knowledge, there are no extensions of the contribution methods of Snidaro and Foresti to 3D fields reported in the literature. Following the definition \cite{3D_EC}, there are 22 unique types of vertex contribution subject to symmetry. All 22 unique contributors are described in Table \ref{tab:3D_bitmap_contributions}. Similarly, the empty type contributes nothing to EC, perimeter, area, or volume, so is only back-calculated in post if desired. This means that with 21 unique contribution types without the empty type, there are a total of 42 array entries per level. Therefore with an 8-bit 3D data file, a storage size of 43176 bytes per vertex contribution array would be required. Again, the size is not overwhelmingly large and therefore a 3D extension of Algorithm \ref{alg:2D_contributions_parallel} is sufficient. It should be noted that many 3D fields, especially those from computer tomography (CT) scans, can use much higher resolution (e.g., 14-bit or 16-bit integers). The size of a 512x512x512 scan with 16-bit data requires almost 24 times the memory of the vertex contribution array.

\begin{table}[!htp]
\renewcommand*{\arraystretch}{1.5}
\footnotesize
\caption{3D vertex contributions to overall EC for the 26-C case along with perimeter, area, and volume contributions.}
\begin{center}
\aboverulesep = 0mm \belowrulesep = 0mm
\begin{tabular}{cccccccccc}
\toprule
	Type 									& 	$f_\text{adj}$	& V (26-C) & E (26-C)	& F (26-C) 	& Vox (26-C) 		& EC (26-C) 	&  Perimeter & Area & Volume	\\ 
\midrule
	\threeDq{0.4}{0}{0}{0}{0}{0}{0}{0}{0} 					&     0 	&     0  	&      0 		&     0 		&        0 	  		&     0 	 	&      0 	 &      0   	& 0 \\
	\threeDq{0.4}{255}{0}{0}{0}{0}{0}{0}{0} 				&     0 	&     1  	&      3 * (0.5) 	&     3 * (0.25) 	&        1 * (0.125)	&     0.125 	&      1.5 	 &      0.75	& 0.125    \\
	\threeDq{0.4}{255}{255}{0}{0}{0}{0}{0}{0} 				&     1 	&     1  	&      4 * (0.5) 	&     5 * (0.25) 	&        2 * (0.125) 	&     0 		&      1 	 &      1 	& 0.25    \\
	\threeDq{0.4}{255}{0}{0}{255}{0}{0}{0}{0} 				&     1.414 	&     1  	&      5 * (0.5) 	&     6 * (0.25) 	&        2 * (0.125) 	&     -0.25	 	&      3 	 &      1.5 	& 0.25    \\
	\threeDq{0.4}{255}{0}{0}{0}{0}{0}{0}{255} 				&     1. 732	&     1  	&      6 * (0.5) 	&     6 * (0.25) 	&        2 * (0.125) 	&     -0.75 		&      3 	 &      1.5 	& 0.25    \\
	\threeDq{0.4}{255}{255}{255}{0}{0}{0}{0}{0} 			&     3.414 	&     1  	&      5 * (0.5) 	&     7 * (0.25) 	&        3 * (0.125) 	&     -0.125 	&      1.5 	 &      1.25	& 0.375    \\
	\threeDq{0.4}{255}{255}{0}{0}{0}{0}{255}{0} 			&     4.146 	&     1  	&      6 * (0.5) 	&     8 * (0.25) 	&        3 * (0.125) 	&     -0.375	&      2.5 	 &      1.75	& 0.375    \\
	\threeDq{0.4}{255}{0}{0}{255}{0}{0}{255}{0} 			&     4.243		&     1  	&      6 * (0.5) 	&     9 * (0.25) 	&        3 * (0.125) 	&     -0.125 	&      4.5 	 &      2.25	& 0.375    \\
	\threeDq{0.4}{255}{255}{255}{255}{0}{0}{0}{0} 			&     6.828		&     1  	&      5 * (0.5) 	&     8 * (0.25) 	&        4 * (0.125) 	&     0 	 	&      0 	 &      1 	& 0.5    \\
	\threeDq{0.4}{255}{255}{255}{0}{255}{0}{0}{0} 			&     7.243		&     1  	&      6 * (0.5) 	&     9 * (0.25) 	&        4 * (0.125) 	&     -0.25	 	&      3 	 &      1.75	& 0.5    \\
	\threeDq{0.4}{255}{255}{255}{0}{0}{255}{0}{0} 			&     7.560		&     1  	&      6 * (0.5) 	&     9 * (0.25) 	&        4 * (0.125) 	&     -0.25 	 	&      2 	 &      1.5 	& 0.5    \\
	\threeDq{0.4}{255}{255}{255}{0}{0}{0}{0}{255} 			&     7.975		&     1  	&      6 * (0.5) 	&     10 * (0.25) 	&        4 * (0.125) 	&     0 	 	&      3 	 &      2 	& 0.5    \\
	\threeDq{0.4}{255}{0}{255}{0}{0}{255}{0}{255} 			&     8.293		&     1  	&      6 * (0.5) 	&     10 * (0.25) 	&        4 * (0.125) 	&     0	 	&      2 	 &      2 	& 0.5    \\
	\threeDq{0.4}{255}{0}{0}{255}{0}{255}{255}{0} 			&     8.485		&     1  	&      6 * (0.5) 	&     12 * (0.25) 	&        4 * (0.125) 	&     0.5 	 	&      6 	 &      3 	& 0.5    \\
	\threeDq{0.4}{255}{0}{255}{0}{255}{0}{255}{255} 		&     3.414$^\dag$	&     1  	&      6 * (0.5) 	&     10 * (0.25) 	&        5 * (0.125) 	&     -0.125 	&      1.5 	 &      1.25	& 0.625    \\
	\threeDq{0.4}{255}{0}{255}{0}{0}{255}{255}{255}  		&     4.146$^\dag$	&     1  	&      6 * (0.5) 	&     11 * (0.25) 	&        5 * (0.125) 	&     0.125 	&      2.5 	 &      1.75	& 0.625    \\
	\threeDq{0.4}{255}{0}{0}{255}{255}{255}{255}{0} 		&     4.243$^\dag$	&     1  	&      6 * (0.5) 	&     12 * (0.25) 	&        5 * (0.125) 	&     0.375 	&      4.5 	 &      2.25	& 0.625    \\
	\threeDq{0.4}{255}{0}{255}{0}{255}{255}{255}{255} 		&     1$^\dag$	&     1  	&      6 * (0.5) 	&     11 * (0.25) 	&        6 * (0.125) 	&     0 	 	&      1 	 &      1 	& 0.75    \\
	\threeDq{0.4}{255}{0}{0}{255}{255}{255}{255}{255} 		&     1.414$^\dag$	&     1  	&      6 * (0.5) 	&     12 * (0.25) 	&        6 * (0.125) 	&     0.25 	 	&      3 	 &      1.5 	& 0.75    \\
	\threeDq{0.4}{255}{255}{255}{0}{0}{255}{255}{255} 		&     1.732$^\dag$	&     1  	&      6 * (0.5) 	&     12 * (0.25) 	&        6 * (0.125) 	&     0.25 	 	&      3 	 &      1.5 	& 0.75    \\
	\threeDq{0.4}{255}{255}{255}{255}{255}{255}{255}{0} 	&     0$^\dag$	&     1  	&      6 * (0.5) 	&     12 * (0.25) 	&        7 * (0.125) 	&     0.125 	&      1.5 	 &      0.75	& 0.875    \\
	\threeDq{0.4}{255}{255}{255}{255}{255}{255}{255}{255} 	&     0$^\dag$	&     1  	&      6 * (0.5) 	&     12 * (0.25) 	&        8 * (0.125) 	&     0 	 	&      0 	 &      0 	& 1    \\[1.3ex]
\bottomrule
\multicolumn{9}{l}{\small $^\dag$ Adjacency of empty voxels used to avoid more computation}
\end{tabular}
\aboverulesep = 0.605mm \belowrulesep = 0.984mm
\end{center}
\label{tab:3D_bitmap_contributions}
\end{table}

A major difference between the 2D and 3D case is that diagonal adjacency of cells in the vertex neighborhood must be determined to differentiate between contributor types in the 3D case whereas only one adjacency needs to be calculated for the 2D case. Using the sum of Euclidean distance between all pairs of points, one can completely differentiate contribution types and use a running sum to avoid computing the same number twice. We call this function $f_\text{adj}$ and this defined as:
\begin{align}
	f_\text{adj}\left(\mathcal{V}\right) = \sum_{i=1}^{\lvert\mathcal{V}\rvert - 1} \sum_{j=i+1}^{\lvert\mathcal{V}\rvert} \norm{V_i - V_j}_2.
\end{align}
Here, $\mathcal{V}$ represents the set of vertex coordinates $V_i$. The 2-norm is taken to get the Euclidean distance between all coordinate pairs of the vertices, allowing for distinction between which type of contribution is present. For instance, with 2 active cells, they may share only one face ($f_\text{adj} = 1$), share only one edge ($f_\text{adj} = \sqrt{2}$), or share only one vertex ($f_\text{adj} = \sqrt{3}$). Adjacency for all configurations of vertex neighborhoods in 3D are included in Table \ref{tab:3D_bitmap_contributions}. Also, for the case of 5, 6, and 7 cells active, the position of the inactive, or empty, cells is used to reduce the number of adjacencies computed. For the sake of brevity, the serial version for computing 3D vertex contributions is shown in the appendix in Algorithm \ref{alg:3D_contributions_serial}. When considering memory requirements, the implementation of a low-memory algorithm for 3D field processing is all the more important. For this case, the 3D algorithm can be extended exactly the same way as the 2D case, by replacing the calls to the 8-cell values in the field file in memory with calls to read the 8 cell-specific indices from the field file.

\subsection{Low-Memory Processing}

Reading a data file for processing often requires that the entire field is held in memory during execution. To mitigate this, the same method that is utilized for the serial and parallel cases can be implemented by forgoing reading the whole field into memory, and instead reading only the values needed from the file at a given vertex of the field. To compute a vertex contribution in two dimensions, only four values are required to be read into memory at a given time. Exploiting the file structure in the BMP file format (for bitmap images), we utilize a low memory version of Algorithm \ref{alg:2D_contributions_serial} (lines \ref{line:2D_data_start} through \ref{line:2D_data_stop}) and Algorithm \ref{alg:2D_contributions_parallel} (lines \ref{line:2D_data_start_parallel} through \ref{line:2D_data_stop_parallel}) by replacing calls to the Field data (i.e., data$[0] \gets$ Field$(i - 1, j - 1)$) with calls to the raw data file (i.e., data$[0] \gets$ read(Field face$(i - 1, j - 1)$)).

\section{Computational Results}

The proposed algorithms were implemented in C\texttt{++} and invoked from {\tt Python}. For the low-memory cases, C\texttt{++} was also used to read in the partial field data. A comparison between the {\tt GUDHI} software package \cite{GUDHI} called from {\tt Python} and the algorithms for 2D and 3D field analysis shown above was performed and reported in the following sections. The EC was first be computed on synthetic random fields of standard sizes for 2D and 3D fields. We then analyze data arising in real applications: microscopy (2D), molecular dynamics simulations (3D), and hyperspectral images (3D). For all case studies, data was translated from raw format (i.e., float or integer) to integer format using values from 0 to 255. For each case, benefit of parallelization was measured by computing the EC with up to 24 CPU cores. All timing results were obtained on a 24-core (Intel Xeon E5-2697—2.7 GHz) computing server with 256 GB of RAM running Red Hat Enterprise release 6.10. All data and software needed for reproducing the results are shared as open-source code and are available at \url{https://github.com/zavalab/ML/tree/master/FastTopology}. 

\subsection{Analysis of Synthetic Random Fields}

\begin{figure}[!htp]
    \centering
    \begin{subfigure}{0.4\textwidth}
        \centering
        \includegraphics[width=2.5in]{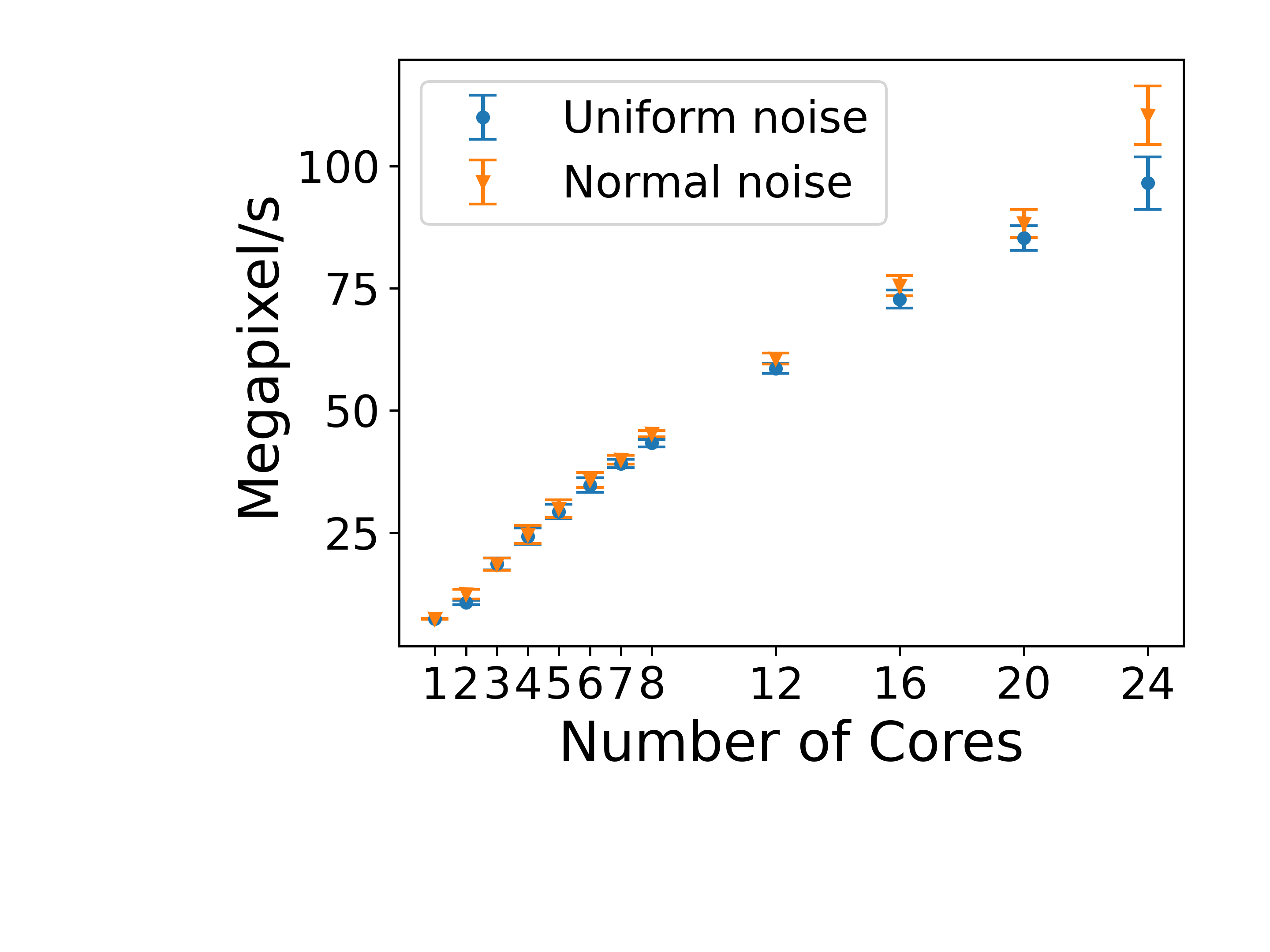}
        \caption{Processing speed}
    \end{subfigure}%
    \hfill
    \begin{subfigure}{0.4\textwidth}
        \centering
        \includegraphics[width=2.5in]{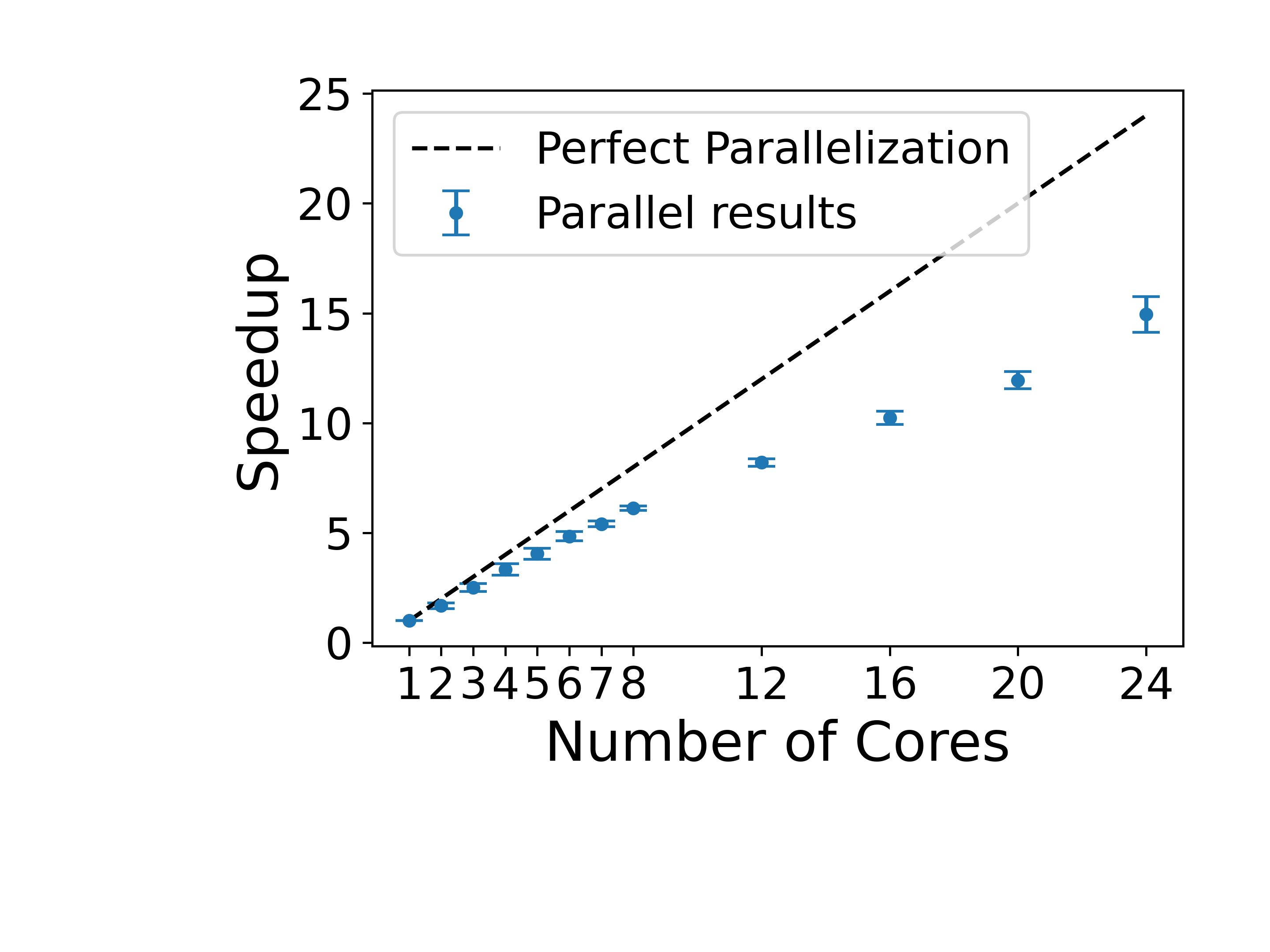}
        \caption{Speedup from parallelization}
    \end{subfigure}
    \hfill
    \caption{Timing results for fields of size 2048x2048. On the left, the processing speed of faces for Algorithm \ref{alg:2D_contributions_parallel} with 2-24 cores. On the right, the speedup due to parallelization of Algorithm \ref{alg:2D_contributions_parallel} with 2-24 cores compared to a perfect parallel efficiency line (dashed).}
    \label{fig:speedup_2048x2048}
\end{figure}

We generated synthetic 2D random fields in order to test the scalability of the proposed algorithms in a systematic manner. Standard 2D image sizes of 1280x720 and 1920x1080 were tested along with a couple of large square field sizes of dimension 2048x2048 and 4096x4096. Each field size was generated using 500 random samples of both uniform and random noise to show run-to-run timing variation. A summary of the results is provided  in Table \ref{tab:2D_random_timings}. We can see that the processing time for computing the EC using the parallel Algorithm \ref{alg:2D_contributions_parallel} is nearly 3 orders of magnitude faster than {\tt GUDHI}, with serial operation already being 50 times faster. The speedup obtained using more cores can be seen in Figure \ref{fig:speedup_2048x2048} for a field of size 2048x2048. On the left we show that the total Megapixels (MP) processed per second exceeds 100 when using 24 cores and the parallel speedup efficiency is 62\%. Processing random fields of similar size using 8 cores with {\tt CHUNKYEuler} results in a processing speed of about 46.5 MP/s as reported in \cite{Wang_EC_GPU}, compared to 43.3 - 45.2 MP/s using Algorithm \ref{alg:2D_contributions_parallel} with 8 cores. As such, our implementation is as scalable as that of {\tt CHUNKYEuler}. 

\begin{table}[!htp]
\footnotesize
\caption{Timing results in seconds and million pixels (faces) processed per second for random 2D fields.}
\begin{center}
\begin{tabular}{ccccc}
\toprule
	Type 	& GUDHI (s) 	& GUDHI (MP/s)	& Alg. \ref{alg:2D_contributions_parallel}, 24 cores (s)	& Alg. \ref{alg:2D_contributions_parallel}, 24 cores (MP/s)	\\ 
\midrule
	1280x720 U  	&     5.31 $\pm$ 0.16 	&      0.174 $\pm$ 0.005 	&     	0.0095 $\pm$ 0.0006  	&       97.7 $\pm$ 4.7 	\\
	1280x720 N	&     5.11 $\pm$ 0.16 	&      0.181 $\pm$ 0.005 	&     	0.0093 $\pm$ 0.0006  	&       99.3 $\pm$ 4.9  	\\
	1920x1080 U 	&     14.99 $\pm$ 0.44 	&      0.138  $\pm$ 0.003	& 	0.0193 $\pm$ 0.0010 	&       107.4 $\pm$ 4.1  	\\
	1920x1080 N  	&     14.57 $\pm$ 0.44 	&      0.142 $\pm$ 0.004 	&      0.0188 $\pm$ 0.0011 	&       110.8 $\pm$ 4.6  	\\
	2048$^2$ U 	&     32.50 $\pm$ 1.24 	&      0.129 $\pm$ 0.004 	&     	0.0436 $\pm$ 0.0031 	&       96.5 $\pm$ 5.3   	\\
	2048$^2$ N  	&     31.35 $\pm$ 1.07 	&      0.133 $\pm$ 0.004 	&     	0.0381 $\pm$ 0.0029 	&       110.4 $\pm$ 6.0   	\\
	4096$^2$ U  	&     151.1 $\pm$ 4.8 	&      0.111 $\pm$ 0.003 	&     0.207 $\pm$ 0.024 	&        82.2 $\pm$ 9.6   \\
	4096$^2$ N	&     145.2 $\pm$ 5.0 	&      0.116 $\pm$ 0.003 	&     0.210 $\pm$ 0.032 	&        81.95 $\pm$ 13.2   \\
\bottomrule
\end{tabular}
\end{center}
\label{tab:2D_random_timings}
\end{table}

\begin{figure}
    \centering
    \begin{subfigure}{0.4\textwidth}
        \centering
        \includegraphics[width=2.5in]{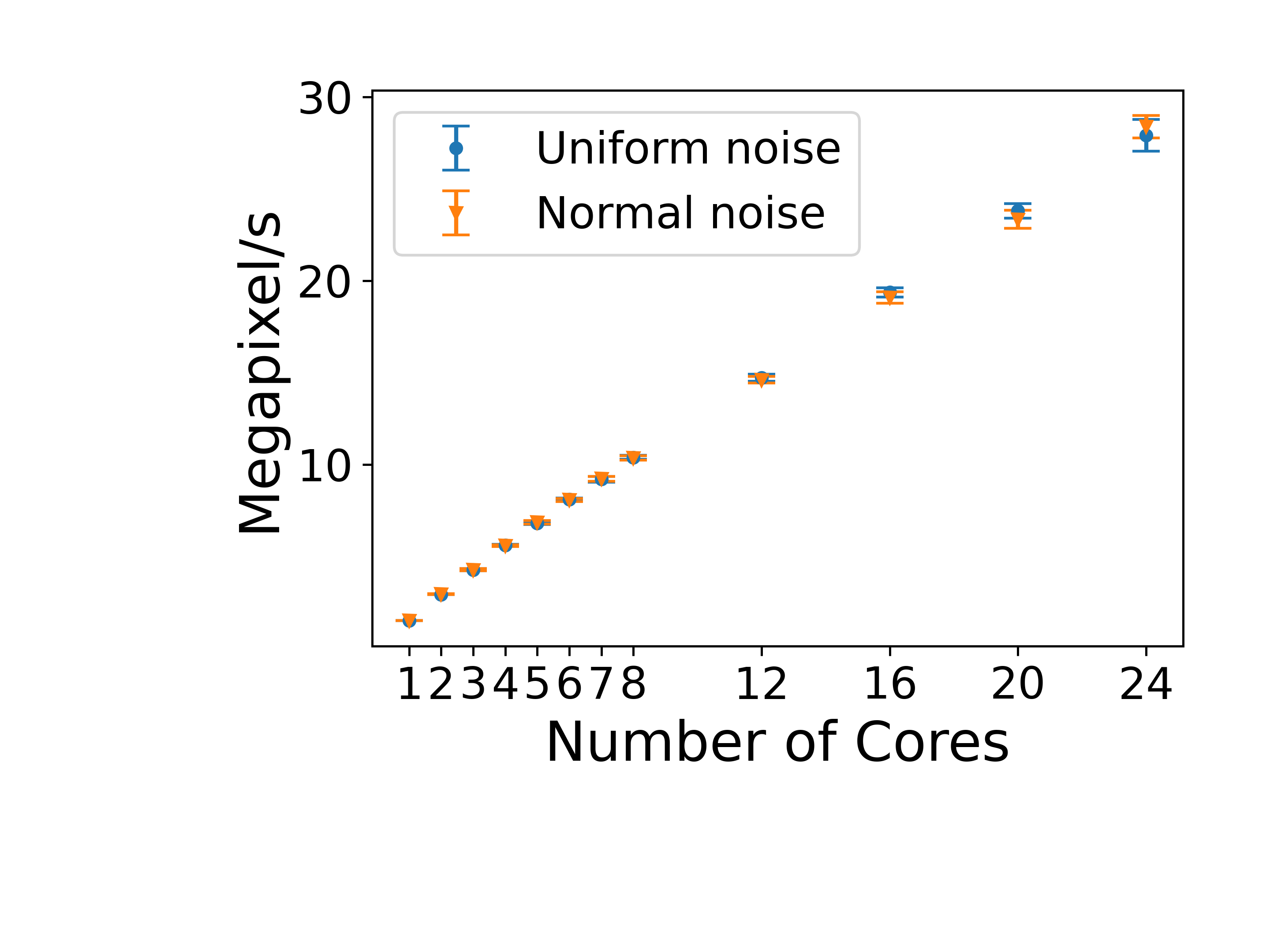}
        \caption{Processing speed}
    \end{subfigure}
    \hfill
    \begin{subfigure}{0.4\textwidth}
        \centering
        \includegraphics[width=2.5in]{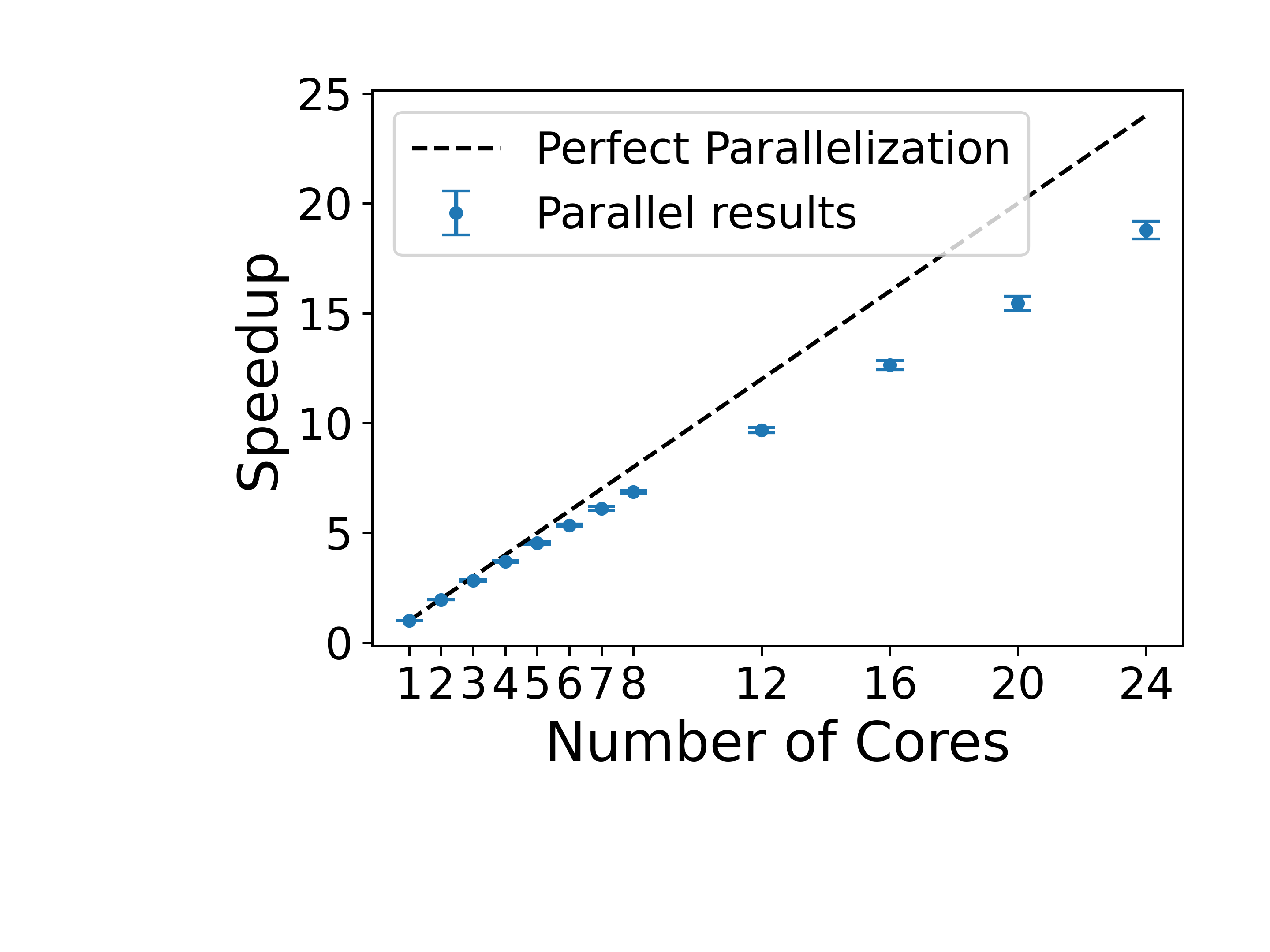}
        \caption{Speedup from parallelization}
    \end{subfigure}
    \hfill
    \caption{Timing results for fields of size 256x256x256. On the left, the processing speed of voxels for Algorithm \ref{alg:2D_contributions_parallel} using the 3D modifications with 2-24 cores. On the right, the speedup due to parallelization of Algorithm \ref{alg:2D_contributions_parallel} with 3D modifications with 2-24 cores compared to a perfect parallel efficiency line (dashed).}
    \label{fig:speedup_256x256x256}
\end{figure}

The 3D algorithm was also tested on both uniform and normal noise for fields of dimensions 128x128x128 and 256x256x256. Given the larger size of the data and processing time required, only 100 samples were run for each case with both uniform and random noise. The processing results are summarized in Table \ref{tab:3D_random_timings}; a similar trend is observed, as {\tt GUDHI} takes nearly 3 orders of magnitude longer than Algorithm \ref{alg:2D_contributions_parallel} using 24 cores. With fields of size 256x256x256, the parallel implementation with 24 cores speeds up execution by ~19 times, leading to an efficiency of 75-80\% (Figure \ref{fig:speedup_256x256x256}). For reference, {\tt CHUNKYEuler} processes voxels/cells at a speed of 26.6 MV/s on 3D random fields with similar size \cite{Wang_EC_GPU}, compared to 10.4 MV/s with Algorithm \ref{alg:2D_contributions_parallel} for 3D, both using 8 CPU cores. The reduction in speed by our implementation is likely due to the higher number of comparisons and computation required to determine adjacency for vertex neighborhood classification. However, this reduction in speed comes with the benefit of more flexibility of enabling computation of alternative topological descriptors. 

\begin{table}[H]
\footnotesize
\caption{Timing results in seconds and million voxels/cells processed per second for random 3D fields.}
\begin{center}
\begin{tabular}{ccccc}
\toprule
	Type 	& GUDHI (s) 	& GUDHI (MV/s)	& Alg. \ref{alg:2D_contributions_parallel} (3D), 24 cores (s)	& Alg. \ref{alg:2D_contributions_parallel} (3D), 24 cores (MV/s)	\\ 
\midrule
	128$^3$ U 	&     42.61 $\pm$ 1.19	&      0.0493 $\pm$ 0.0011 	&     0.0795 $\pm$ 0.0067 	&        26.67 $\pm$ 1.67   \\
	128$^3$ N  	&     42.13 $\pm$ 1.46 	&      0.0498 $\pm$ 0.0015	&     0.0765 $\pm$ 0.0045  	&        27.47 $\pm$ 1.21   \\
	256$^3$ U 	&     472.3 $\pm$ 15.5 	&      0.0356 $\pm$ 0.0011 	&     0.602 $\pm$ 0.020 	&        27.90 $\pm$ 0.87   \\
	256$^3$ N  	&     470.7 $\pm$ 12.2 	&      0.0357 $\pm$ 0.0009 	&     0.591 $\pm$ 0.013 	&        28.38 $\pm$ 0.61   \\
\bottomrule
\end{tabular}
\end{center}
\label{tab:3D_random_timings}
\end{table}

\subsection{Memory analysis}

The low-memory version of Algorithms \ref{alg:2D_contributions_serial} and \ref{alg:2D_contributions_parallel} were also tested on random 2D fields. First, random matrices of uniform and normal distribution were generated for fields of size 1920x1080. Then, the fields were saved as bitmap image files (an easily readable, raw format). Finally, the C\texttt{++} algorithm directly read and computed the EC for each field. 500 fields were generated with statistics on the processing speed reported in Table \ref{tab:2D_low_mem_timings}. There is a drastic processing speed loss from reading individual pixel values to populate only the 4 values from each vertex neighborhood during computation of the EC curve. However, even the serial methods are still faster than {\tt GUDHI} (see Table \ref{tab:2D_random_timings}) and now require the minimal memory to compute the EC curve. Also, data beyond the size of working memory can be analyzed given that this is a streaming approach, similar to that used in {\tt CHUNKYEuler} \cite{Heiss_EC}. In addition, only 2D fields were explored here and memory management issues are even more important in the 3D case. Similar slowdown should be expected when, if not slightly more given the increase in values analyzed from 4 faces/pixels to 8 cells/voxels.

\begin{table}[H]
\footnotesize
\caption{Timing results in million pixels processed per second for the low memory processing algorithm.}
\begin{center}
\begin{tabular}{cccc}
\toprule
	Type 	& Alg. \ref{alg:2D_contributions_serial} (MP/s)	& Alg. \ref{alg:2D_contributions_parallel}, 12 cores (MP/s) & Alg. \ref{alg:2D_contributions_parallel}, 24 cores (MP/s)	\\ 
\midrule
	1920x1080 U 	&     0.346 $\pm$ 0.005	&      3.040 $\pm$ 0.045 	&     4.285 $\pm$ 0.036   \\
	1920x1080 N 	&     0.347 $\pm$ 0.005	&      2.991 $\pm$ 0.029 	&     4.159 $\pm$ 0.047   \\
\bottomrule
\end{tabular}
\end{center}
\label{tab:2D_low_mem_timings}
\end{table}

\subsection{Microscopy Case Study}

\begin{figure}[!htp]
    \centering
    \includegraphics[width=6in]{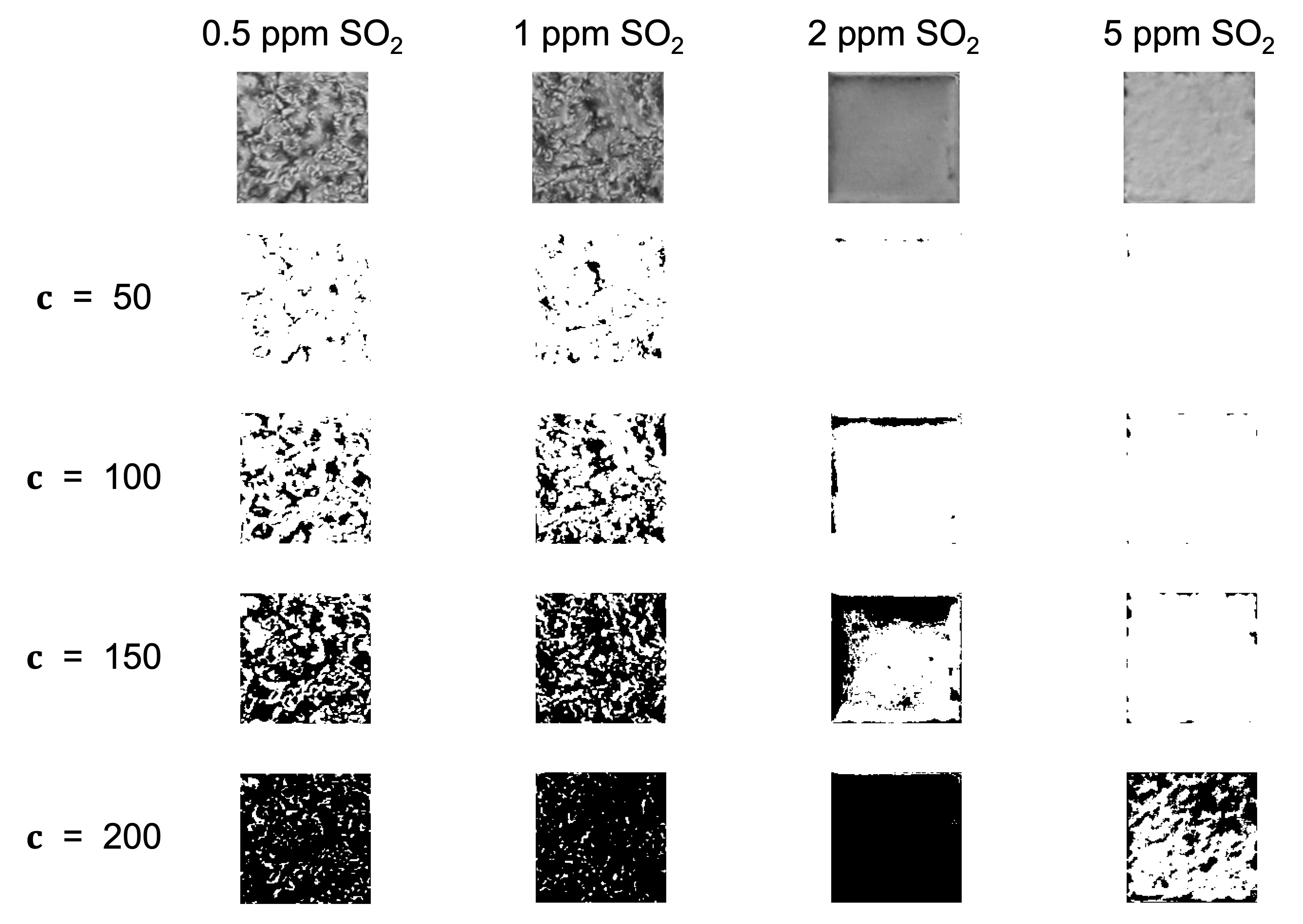}\caption{\mynote{Filtration process for various liquid crystal micrographs. For some micrographs, no activity occurs until over a value of 100, but others activity begins at low filtration values.}}
    \label{fig:SO2_LC_filtrations}
\end{figure}

Liquid crystal sensors elicit optical responses in the presence of target compounds or contaminants. The concentration and environment in which the target compound binds to the liquid has significant impact on the topological state of the system. In this case study, data presented in Jiang et al. \cite{Jiang_LC_TDA} was used to benchmark real data and explore the potential of real-time analysis. The system we are showcasing here is analysis on a gas-based liquid crystal detection system with varying sulfur dioxide (SO$_2$) concentrations, from 0.5 ppm to 5 ppm, in an environment with 40\% relative humidity. An example of how filtration values impact the topology of the system is shown in Figure \ref{fig:SO2_LC_filtrations}. 

\begin{figure}[!htp]
    \centering
    \begin{subfigure}{0.2\textwidth}
        \centering
        \includegraphics[width=1.0in]{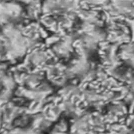}
    \end{subfigure}
    \hfill
    \begin{subfigure}{0.2\textwidth}
        \centering
        \includegraphics[width=1.0in]{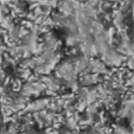}
    \end{subfigure}
    \hfill
    \begin{subfigure}{0.2\textwidth}
        \centering
        \includegraphics[width=1.0in]{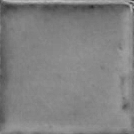}
    \end{subfigure}
    \hfill
    \begin{subfigure}{0.2\textwidth}
        \centering
        \includegraphics[width=1.0in]{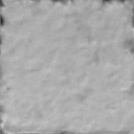}
    \end{subfigure}
    \hfill
    \begin{subfigure}{0.2\textwidth}
        \centering
        \includegraphics[width=1.0in]{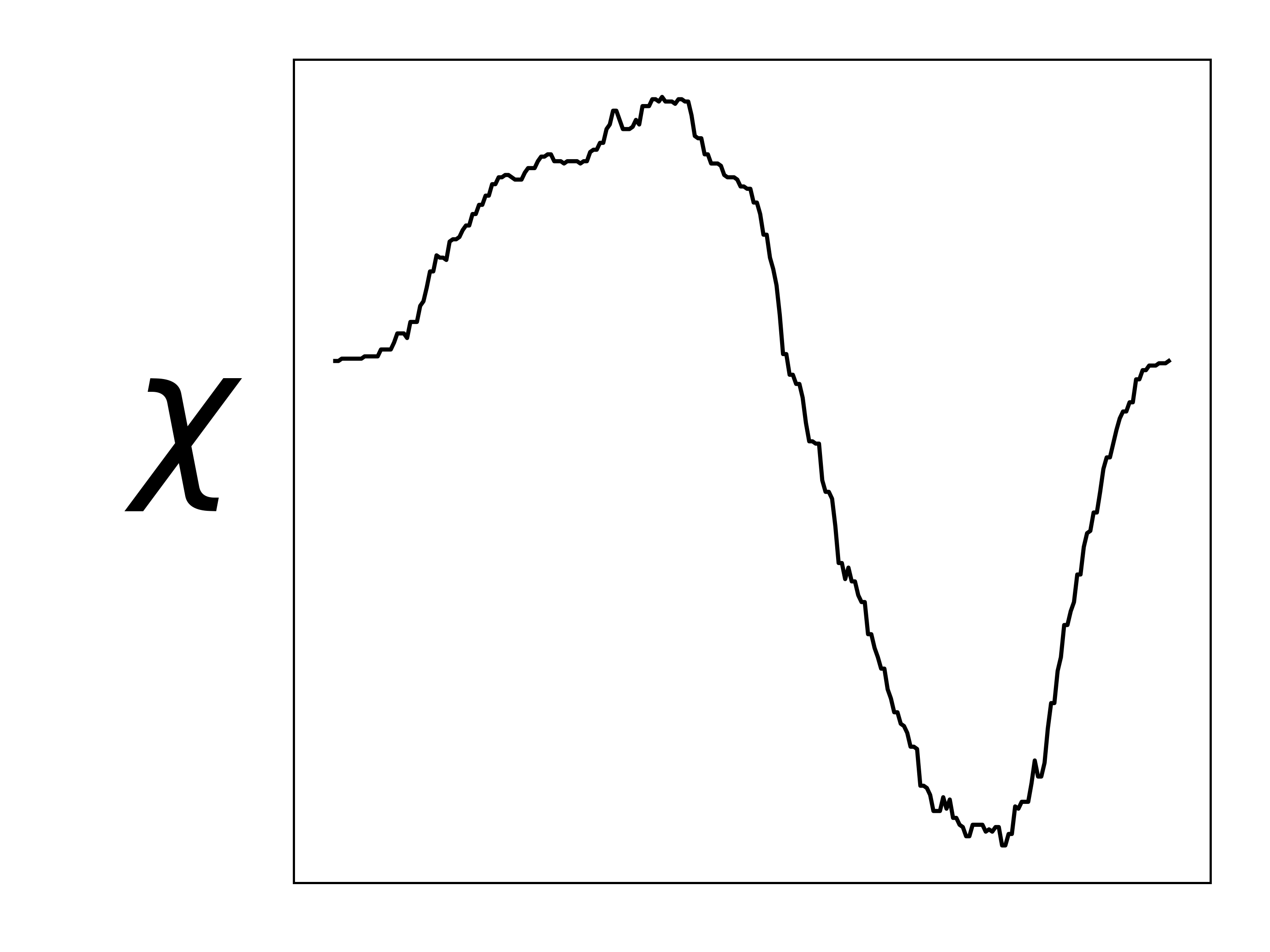}
        \caption{0.5 ppm SO$_2$}
    \end{subfigure}
    \hfill
    \begin{subfigure}{0.2\textwidth}
        \centering
        \includegraphics[width=1.0in]{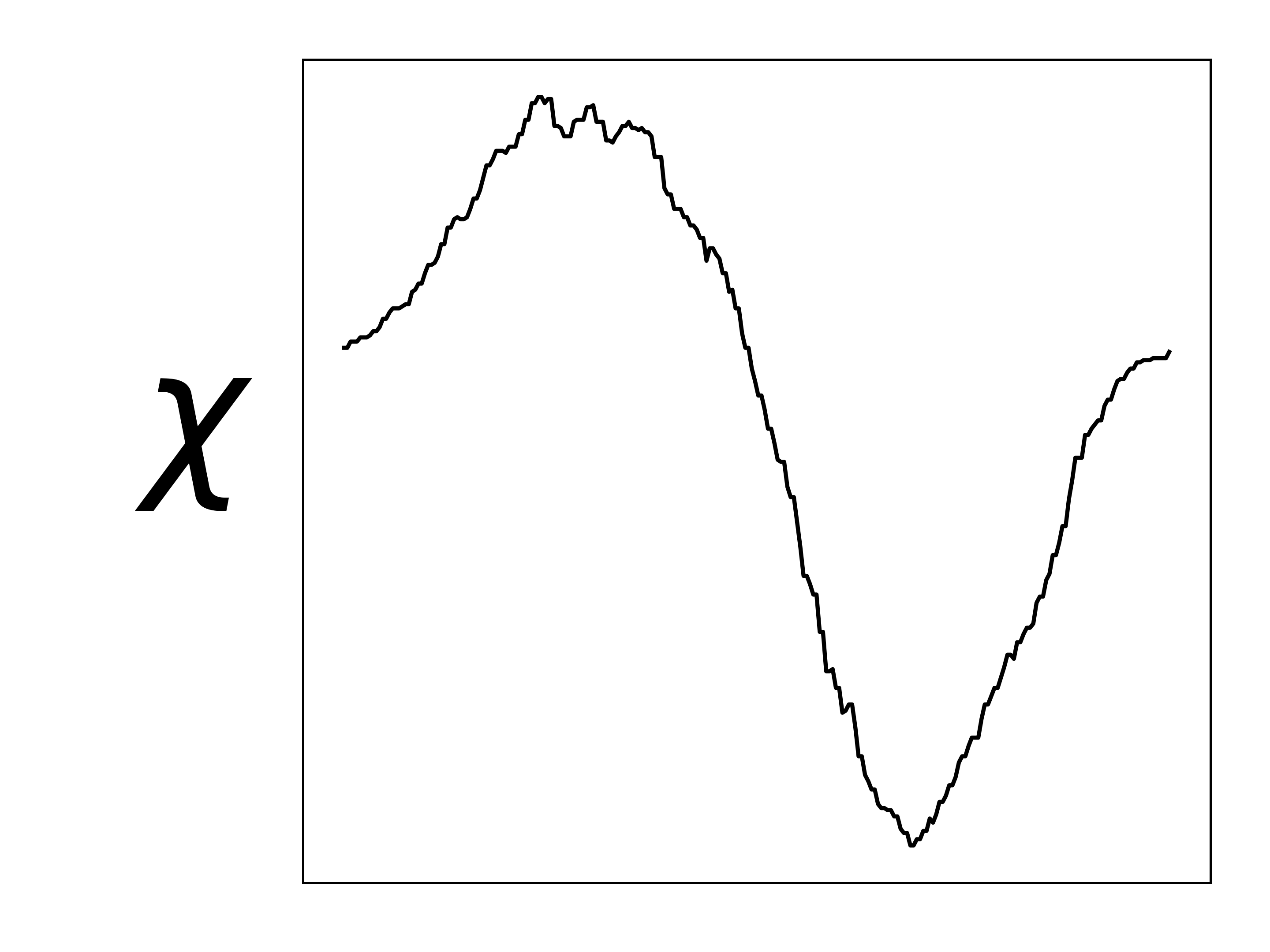}
        \caption{1 ppm SO$_2$}
    \end{subfigure}
    \hfill
    \begin{subfigure}{0.2\textwidth}
        \centering
        \includegraphics[width=1.0in]{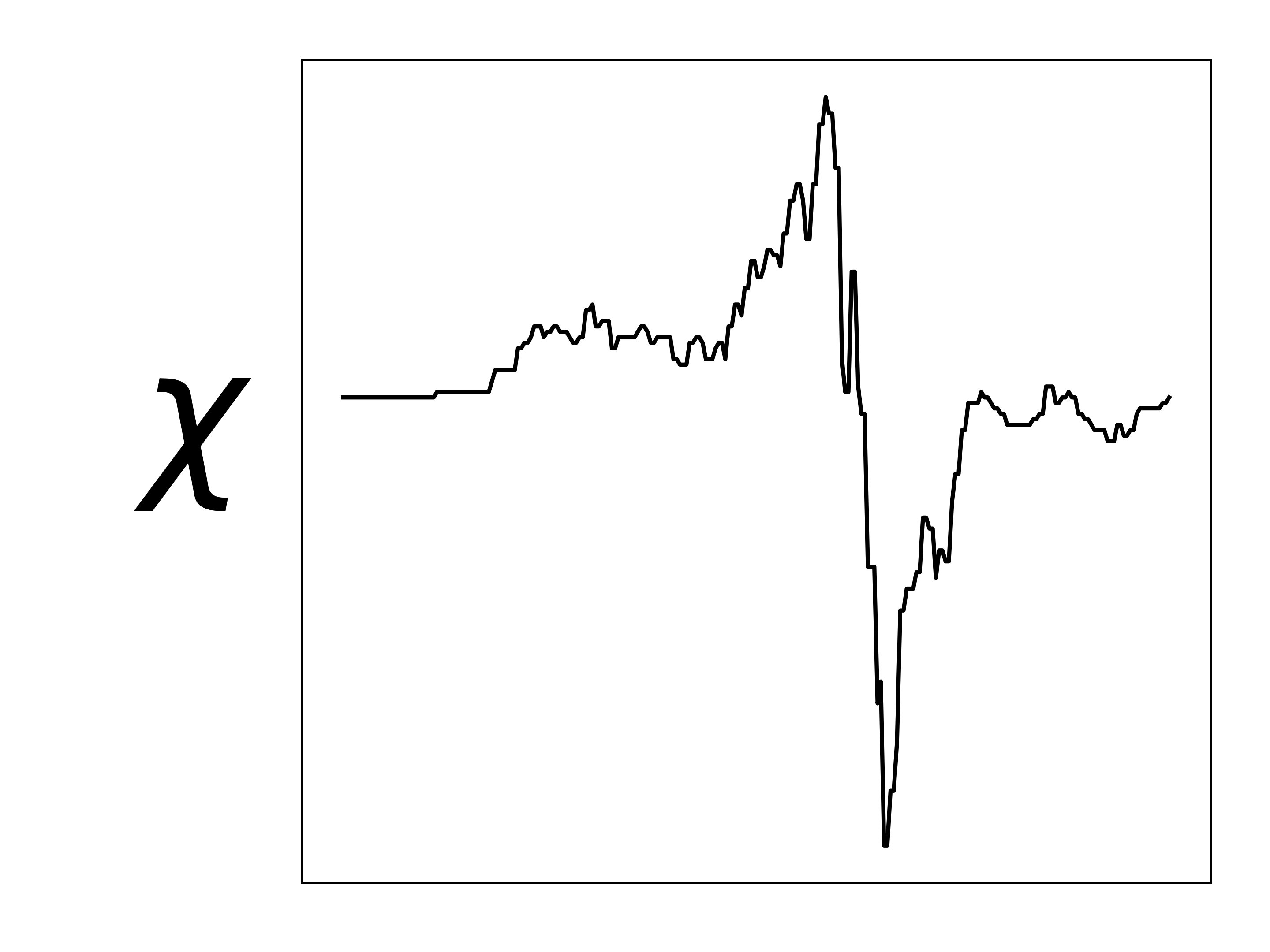}
        \caption{2 ppm SO$_2$}
    \end{subfigure}
    \hfill
    \begin{subfigure}{0.2\textwidth}
        \centering
        \includegraphics[width=1.0in]{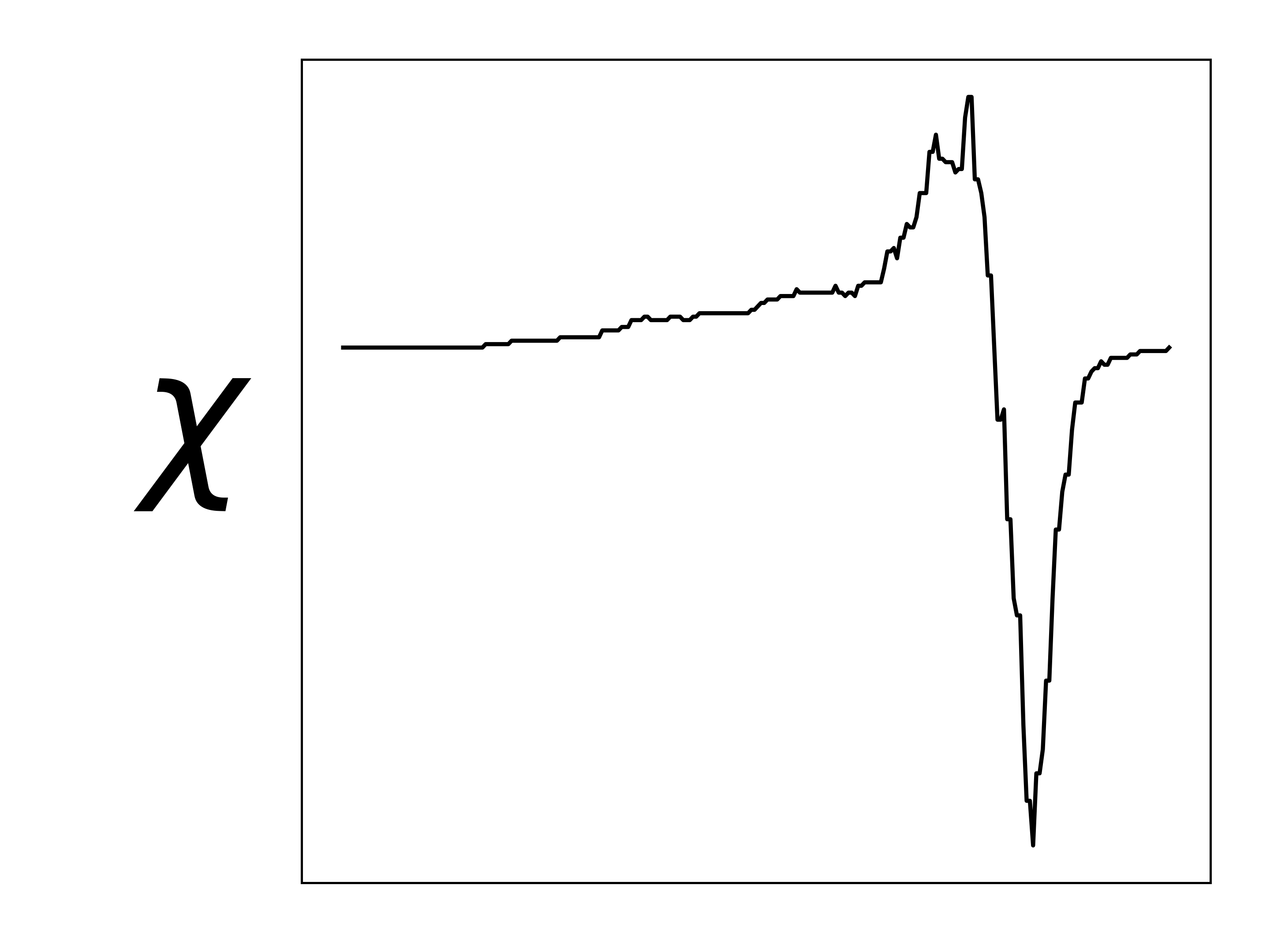}
        \caption{5 ppm SO$_2$}
    \end{subfigure}
    \hfill
    \caption{Varying concentration of SO$_2$ present in the system changes both the optical response and the topological description of the image through the EC. The optical response and subsequent EC curve are shown for 0.5 ppm, 1 ppm, 2 ppm, and 5 ppm SO$_2$ systems in 40\% relative humidity.}
    \label{fig:SO2_LC_states}
\end{figure}

As shown in Figure \ref{fig:SO2_LC_states}, the optical responses at differing SO$_2$ concentrations have distinct EC curves. On a liquid crystal micrograph plate, anywhere from 1 to 36 fully readable grid squares may be used to predict the state of the system. Therefore, the speed in which the images are read dictate how close to real time the sensor can be monitored.

On a single liquid crystal grid, anywhere from 1 to 36 readable grid-squares may be used to predict the state of the system. Therefore, the speed in which the grid-squares from an image or video frame of the sensor are processed dictate how close to real time the sensor can be monitored. In Table \ref{tab:LC_timings}, with image sizes that are so small (approximately 134x134 for each grid-square), parallelization does not receive the same benefit as the larger field data as explored above. There is even a performance decrease by using more than 12 cores for parallel computation; however, the data is processed about 20-30 times faster than {\tt GUDHI}.

\begin{table}[H]
\footnotesize
\caption{Timing results in million pixels processed per second for the liquid crystal sensor case study.}
\begin{center}
\begin{tabular}{ccccc}
\toprule
	Type 	&  GUDHI (MP/s)	& Alg. \ref{alg:2D_contributions_serial} (MP/s)	& Alg. \ref{alg:2D_contributions_parallel}, 12 cores (MP/s) & Alg. \ref{alg:2D_contributions_parallel}, 24 cores (MP/s)	\\ 
\midrule
	SO$_2$ Images 	&     0.332 $\pm$ 0.009	&      6.268 $\pm$ 0.171 	&     11.33 $\pm$ 0.49 &     9.449 $\pm$ 0.225  \\
	Random Fields 	&     0.271 $\pm$ 0.009	&      6.127 $\pm$ 0.132 	&     11.20 $\pm$ 0.39 &     9.600 $\pm$ 0.223    \\
\bottomrule
\end{tabular}
\end{center}
\label{tab:LC_timings}
\end{table}

\subsection{Molecular Dynamics Data}

\begin{figure}[!htp]
    \centering
    \begin{subfigure}{0.3\textwidth}
        \centering
        \includegraphics[width=1.5in]{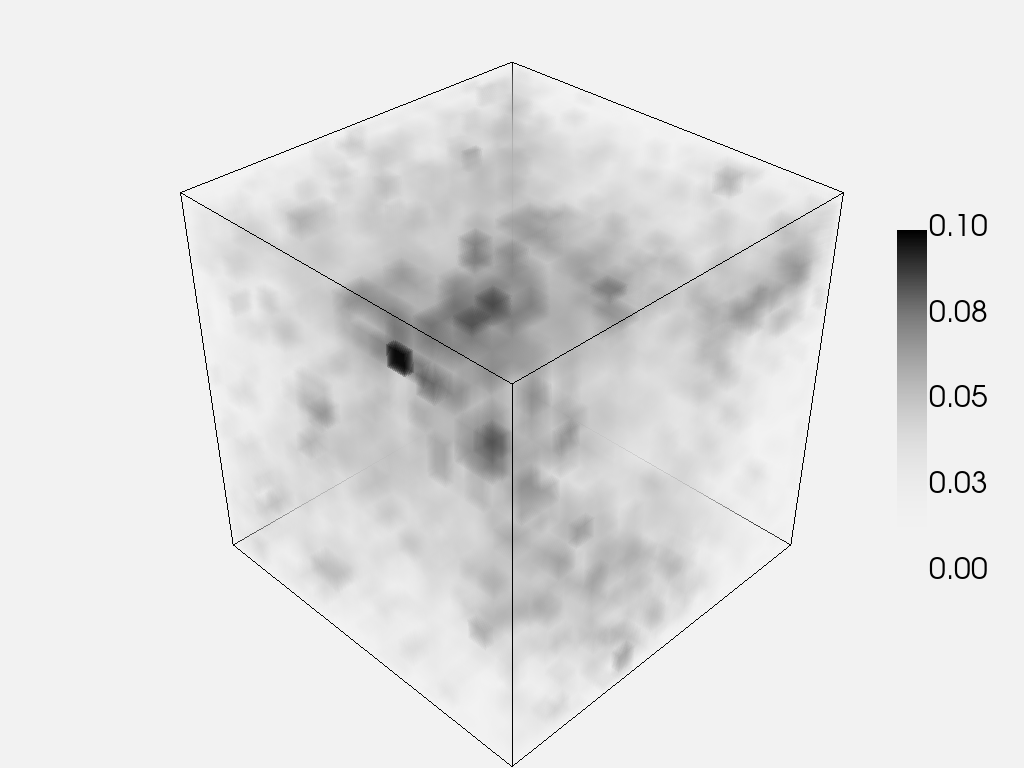}
        \caption{dioxane}
    \end{subfigure}
    \hfill
    \begin{subfigure}{0.3\textwidth}
        \centering
        \includegraphics[width=1.5in]{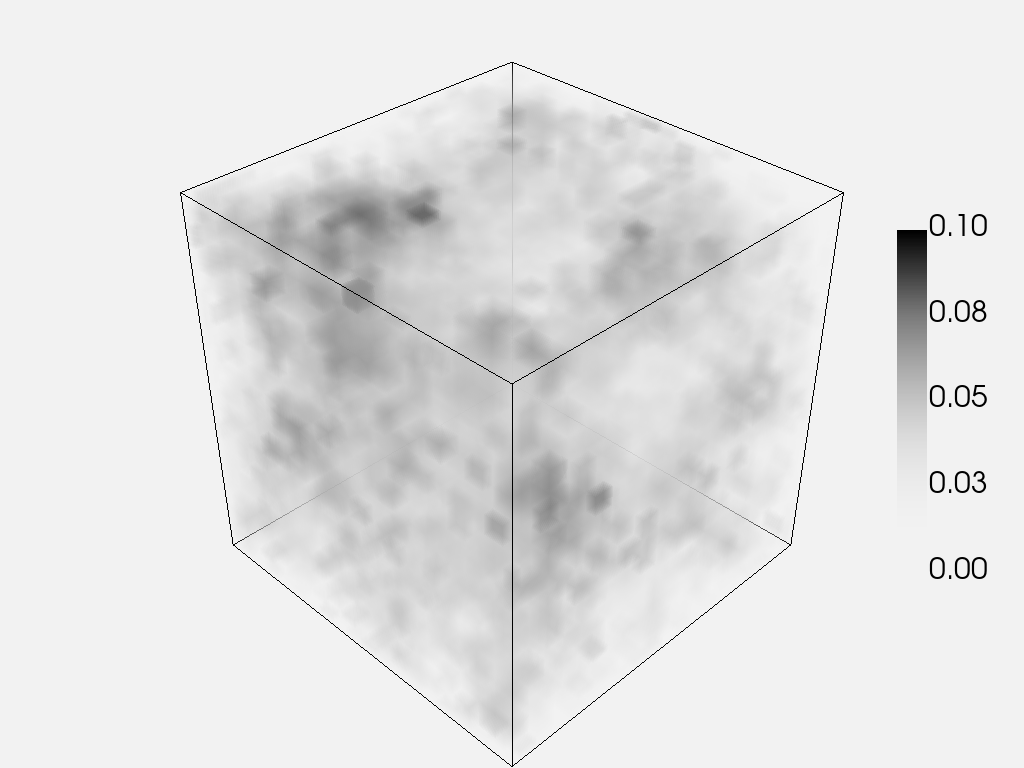}
        \caption{$\gamma$-valerolactone}
    \end{subfigure}
    \hfill
    \begin{subfigure}{0.3\textwidth}
        \centering
        \includegraphics[width=1.5in]{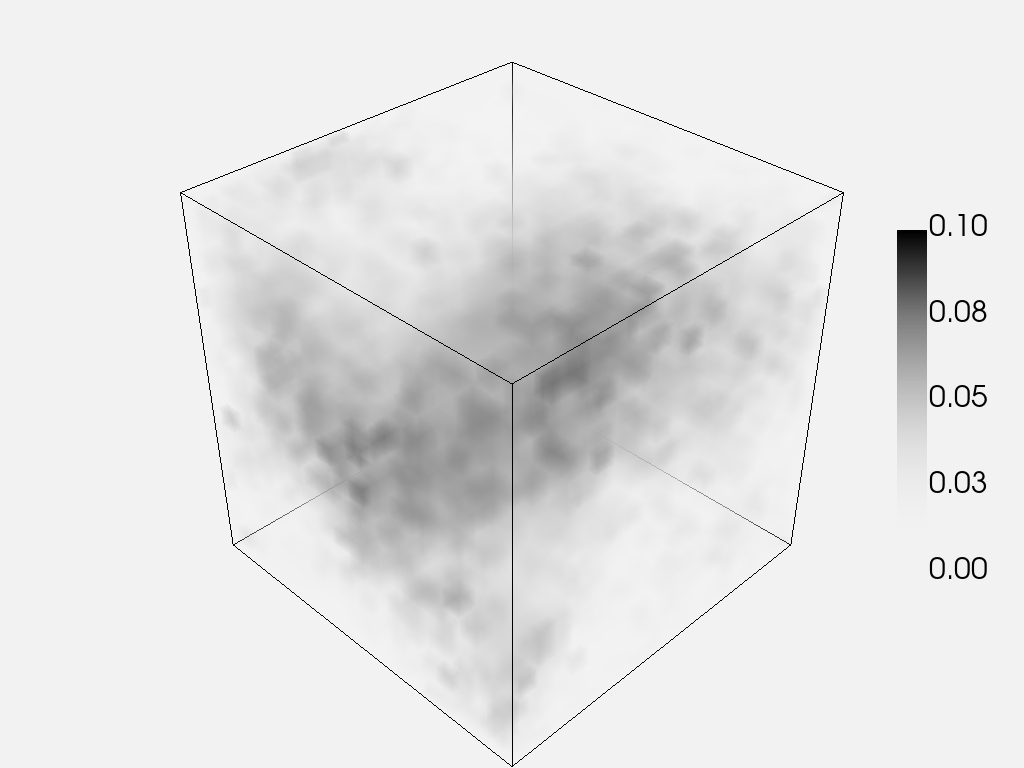}
        \caption{tetrahydrofuran}
    \end{subfigure}
    \hfill
    \caption{Water densities from simulations of fructose with various cosolvent species listed.}
    \label{fig:MD_data}
\end{figure}

\begin{figure}[!htp]
\begin{center}
\includegraphics[width=3in]{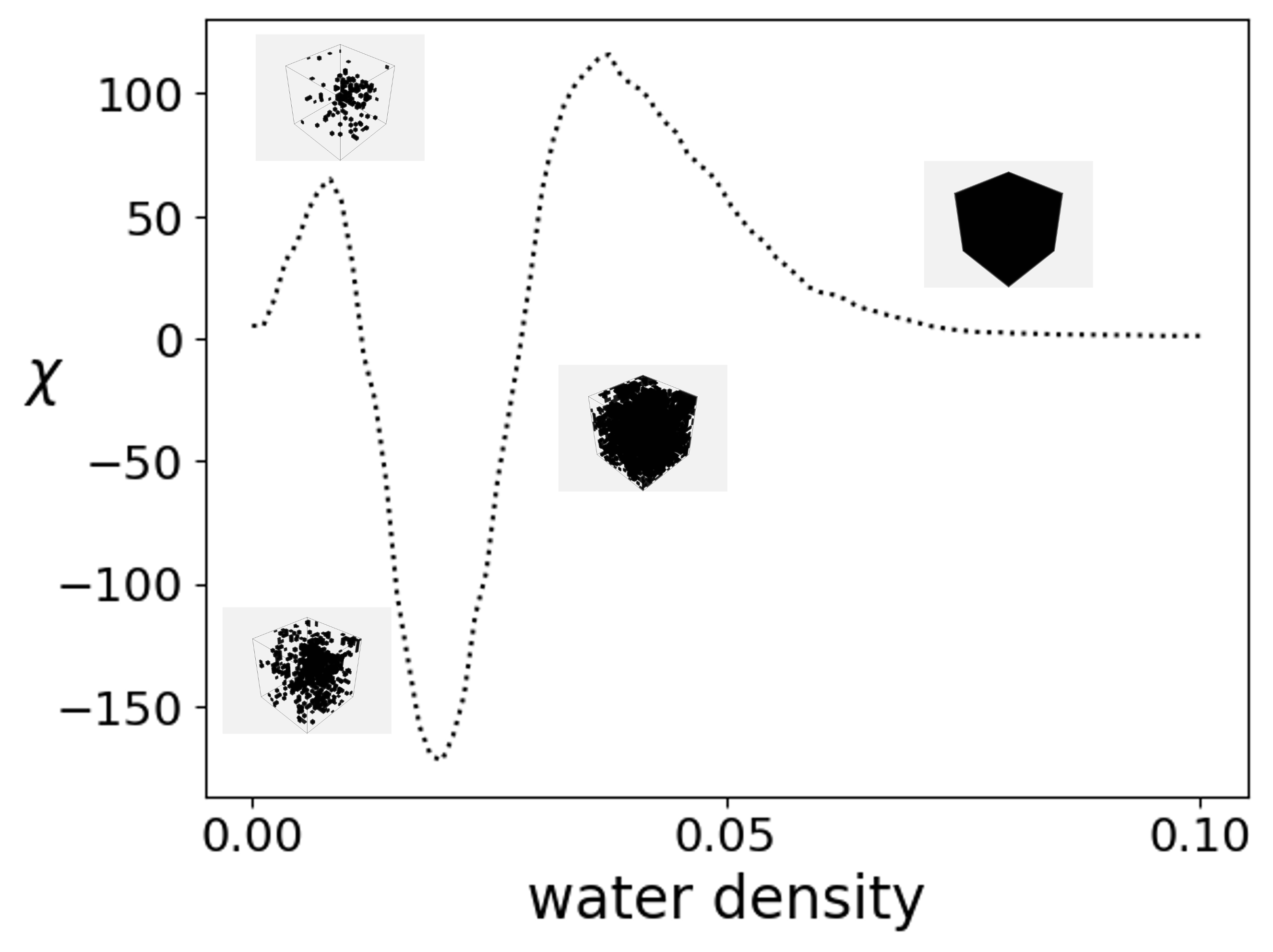}\caption{\mynote{EC curve evolution with 3D binary fields at various filtration values.}}\label{fig:MD_EC_states}
\end{center}
\end{figure}

Simulations that result in 3D fields are also relevant to the Algorithms provided in this work. One such is example is molecular dynamics simulations, where chemical systems can be simulated to understand molecular-scale interactions between participating species. Here, we analyze a system shown by Chew and co-workers  \cite{chew_MD}, and later with the EC by Smith and co-workers \cite{Smith_MD_EC}. The simulations capture  molecular interactions of biomass reactants in cosolvent/water mixtures. As an example, Figure \ref{fig:MD_data} shows the density of water molecules in a 20x20x20 grid for fructose in three different cosolvent/water systems with 10 weight \% water. The topology of the solvent environment has been shown to correlate strongly with reactivity. The process of filtering one of these fields is shown in Figure \ref{fig:MD_EC_states}.

\begin{table}[H]
\footnotesize
\caption{Timing results in million cells processed per second for the molecular dynamics case study.}
\begin{center}
\begin{tabular}{ccccc}
\toprule
	Type 	&  GUDHI (MC/s)	& Alg. \ref{alg:2D_contributions_serial} (3D) (MC/s)	& Alg. \ref{alg:2D_contributions_parallel} (3D), 12 cores (MC/s) & Alg. \ref{alg:2D_contributions_parallel} (3D), 12 cores (MC/s)	\\ 
\midrule
	MD Fields 	&     0.108 $\pm$ 0.004	&      1.065 $\pm$ 0.025 	&     2.105 $\pm$ 0.111  & 1.853 $\pm$ 0.045\\
\bottomrule
\end{tabular}
\end{center}
\label{tab:3D_MD_timings}
\end{table}

\begin{figure}[!htp]
\begin{center}
\includegraphics[width=3in]{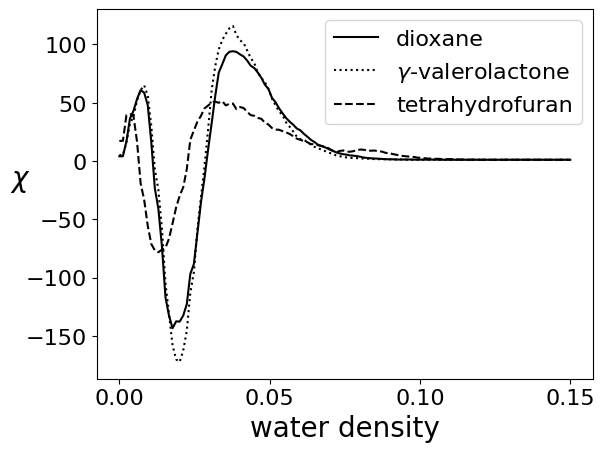}\caption{\mynote{EC Curves for each cosolvent system.}}\label{fig:MD_EC}
\end{center}
\end{figure}

The small size of the data set shows much less scaling advantage for the parallel case, as shown in Table \ref{tab:3D_MD_timings}. Although the timing advantage is less, the proposed algorithm is still one order of magnitude faster than {\tt GUDHI}. To process the entire data set, {\tt GUDHI} took 56.6 seconds, while the serial algorithm took 5.7 seconds, and 12-core parallel implementation took 2.9 seconds. The EC curve of each solvent system is shown in Figure \ref{fig:MD_EC}.

\subsection{Hyperspectral Image Data}

\begin{figure}[!htp]
\begin{center}
\includegraphics[width=3in]{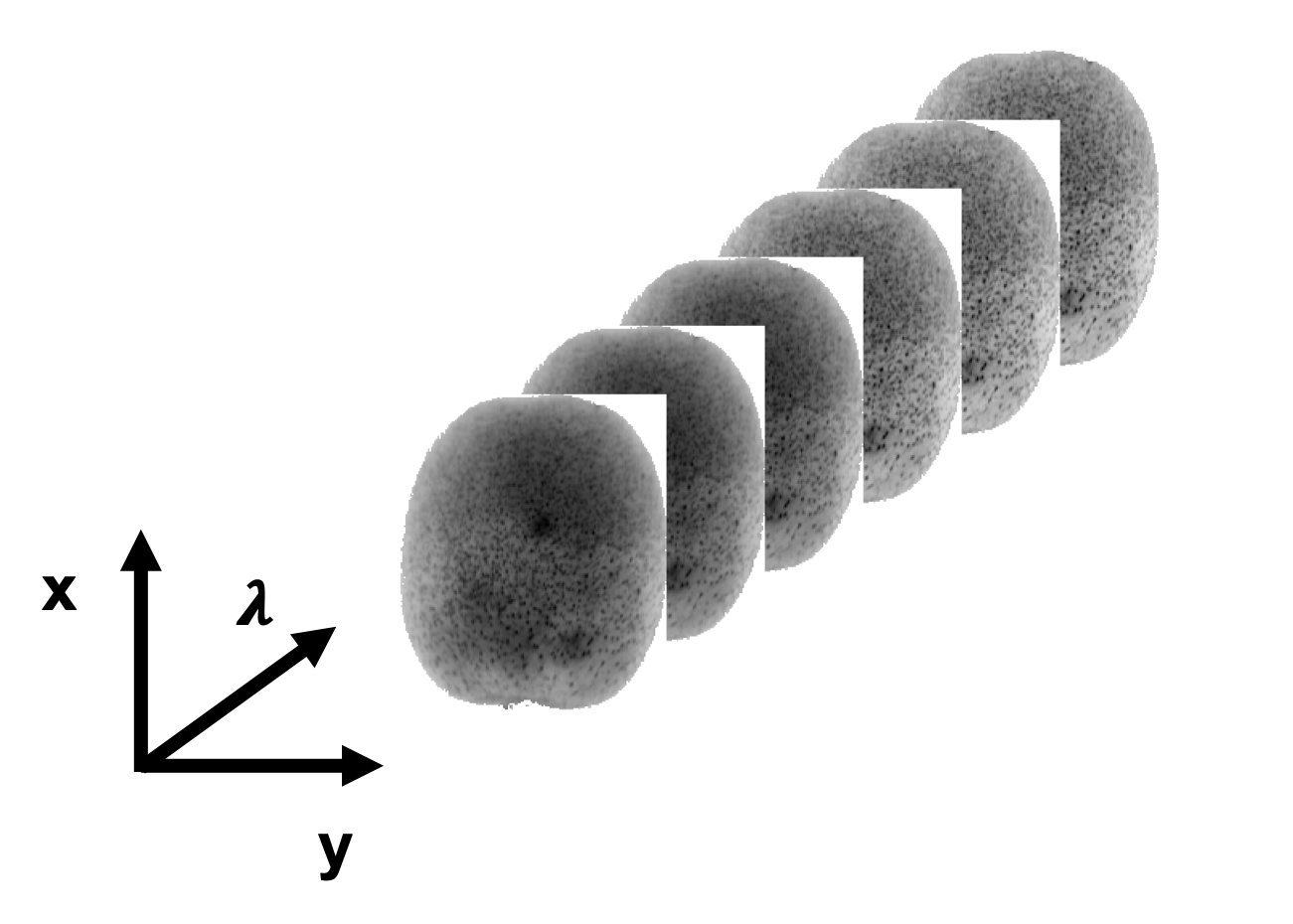}\caption{\mynote{Visualization of hyperspectral image data. Here, a near infrared image with 6 selected wavelengths ($\lambda$) is illustrated. In totality, the hyperspectral image has 252 unique wavelengths.}}\label{fig:HSI_tiled}
\end{center}
\end{figure}

Another example relevant to the chemical sensing community is hyperspectral image analysis. In this research domain, each pixel of a 2D image has a spectral dimension, meaning each pixel of the image is associated with a unique spectra. This requires each image to be analyzed in 3D to take full advantage of the information contained in the spectral dimension. One can view each spectral wavelength as a 2D image, as shown in Figure \ref{fig:HSI_tiled}. 

\begin{figure}[!htp]
    \centering
    \begin{subfigure}{0.3\textwidth}
        \centering
        \includegraphics[width=1.5in]{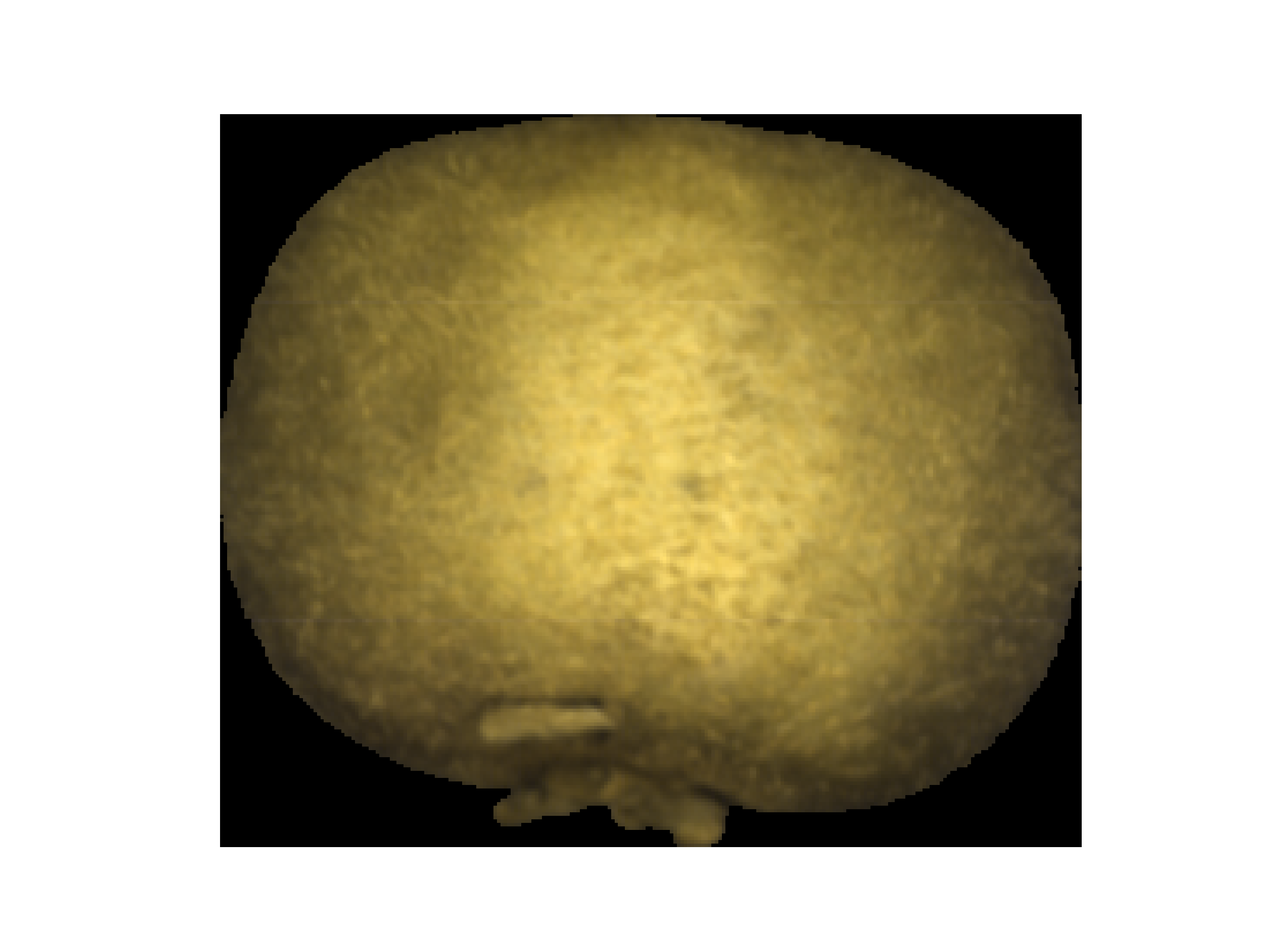}
        \caption{Under ripe}
    \end{subfigure}
    \hfill
    \begin{subfigure}{0.3\textwidth}
        \centering
        \includegraphics[width=1.5in]{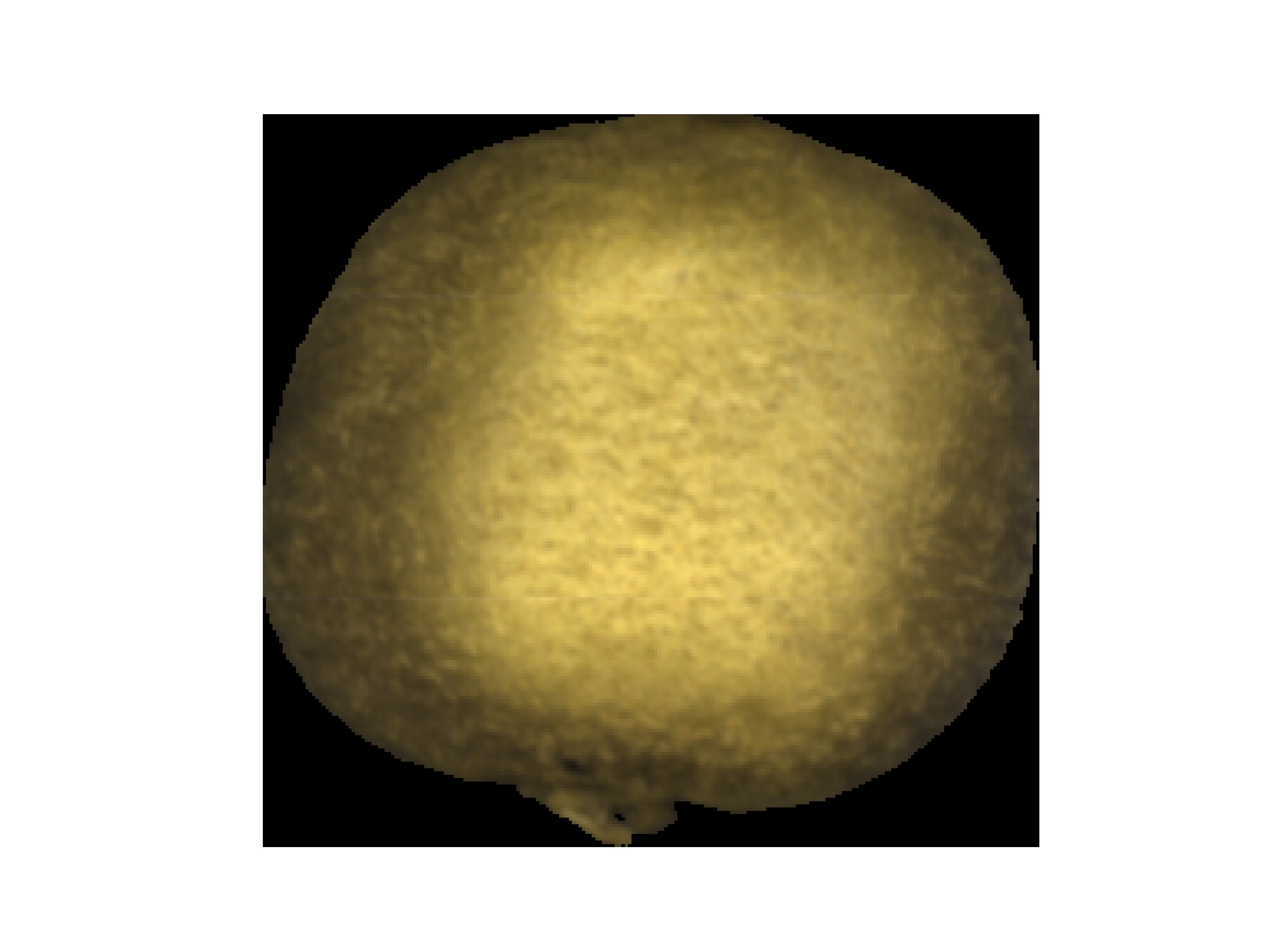}
        \caption{Ripe}
    \end{subfigure}
    \hfill
    \begin{subfigure}{0.3\textwidth}
        \centering
        \includegraphics[width=1.5in]{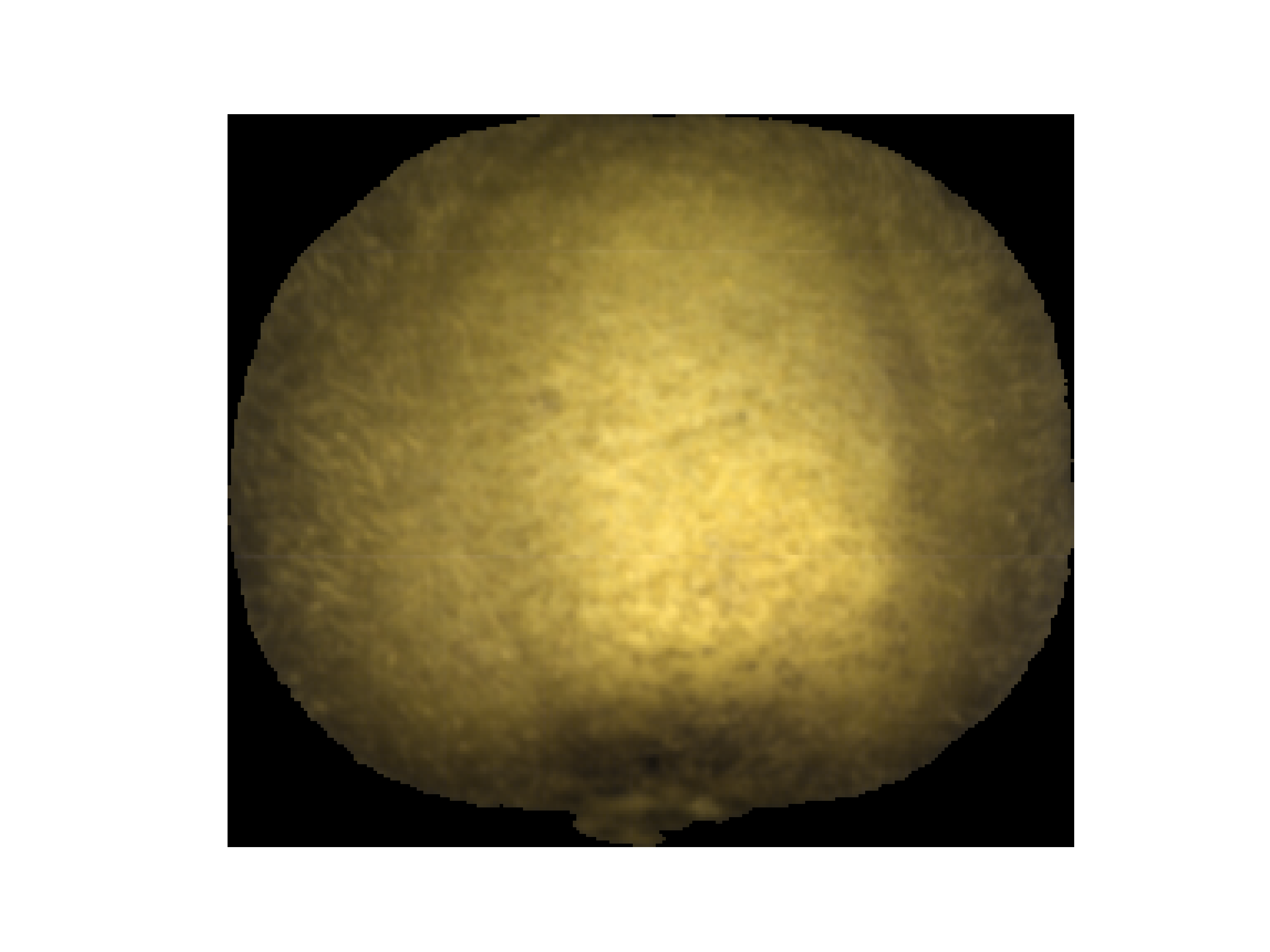}
        \caption{Overripe}
    \end{subfigure}
    \hfill
    \caption{RGB equivalent images of kiwis presenting with varying ripeness levels. RGB alone is difficult to discern whether a kiwi is ripe or not without feeling the firmness of the kiwi.}
    \label{fig:HSI_ripeness}
\end{figure}

A sample data set on the ripeness of fruits \cite{Varga_HSI} is used to benchmark the performance. Hyperspectral images were taken of both the front and back of the fruit each day. Fruits were removed from the set when considered overripe. A total of 360 hyperspectral images make up the kiwi fruit data set. Some sample images translated from visible spectra to RGB are shown in Figure \ref{fig:HSI_ripeness}, where it is clear that using only RGB, it is difficult to tell which kiwi is ripe, overripe, or under ripe. 

\begin{figure}[!htp]
\begin{center}
\includegraphics[width=3in]{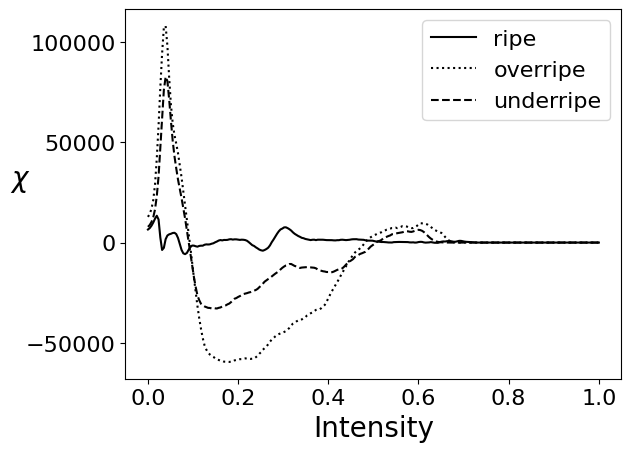}\caption{\mynote{EC Curves for each cosolvent system.}}\label{fig:HSI_EC}
\end{center}
\end{figure}

Statistics on the processing speed are shown in Table \ref{tab:3D_HSI_timings}. The processing time for these images was removed as a column of the table due to the varying size of the images and instead strictly analyzed processing speed in MV/s, similar to the liquid crystal case study. For comparison, the EC curve from filtering the hyperspectral images of ripe, overripe, and under ripe kiwis are shown in Figure \ref{fig:HSI_EC}.

\begin{table}[!htp]
\footnotesize
\caption{Timing results in million voxels processed per second for the hyperspectral imaging case study.}
\begin{center}
\begin{tabular}{cccc}
\toprule
	Type 	&  GUDHI (MV/s)	& Alg. \ref{alg:2D_contributions_serial} (3D) (MV/s)	& Alg. \ref{alg:2D_contributions_parallel} (3D), 24 cores (MV/s)	\\ 
\midrule
	Kiwi Images 	&     0.0701 $\pm$ 0.0078	&      1.973 $\pm$ 0.112 	&     34.57 $\pm$ 2.38  \\
	Random Fields 	&     0.0427 $\pm$ 0.0031	&      1.506 $\pm$ 0.006 	&     28.19 $\pm$ 1.02    \\
\bottomrule
\end{tabular}
\end{center}
\label{tab:3D_HSI_timings}
\end{table}

\section{Discussion}

Using Algorithms \ref{alg:2D_contributions_serial} and \ref{alg:2D_contributions_parallel} in both 2D and 3D field data we find that there is a clear scaling advantage in both time and memory usage while computing EC when compared with off-the-shelf computational tools such as {\tt GUDHI}. Also, when compared to state-of-the-art parallel implementations, such as {\tt CHUNKYEuler}, there is comparable speed. The time complexity of {\tt CHUNKYEuler} and the algorithms presented in this work are nearly identical, as both methods stem from the same principle that using change in features is more computationally scalable than counting features at each filtration value.

For the microscopy case study for liquid crystals, the average size of images in that data set was 134x134, or 17956 pixels, meaning that it would require a processing speed of 19 MP/s to fully process a liquid crystal sensor with 36 readable grid-squares at a video frame rate of 30 frames per second (FPS). Considering control applications, the serial version of the tool is capable of processing 11 grid-squares in real-time considering a 30 FPS video. However, some control applications will not require the reading of all 36 grid-squares, or reading data as quick as 30 measurements per second. Also, when implementing a complete control scheme, the time required to read and crop the raw video data to these processable grid-squares must be considered, introducing a time delay between measurement and data reconciliation.

\begin{figure}[!htp]
\begin{center}
\includegraphics[width=3in]{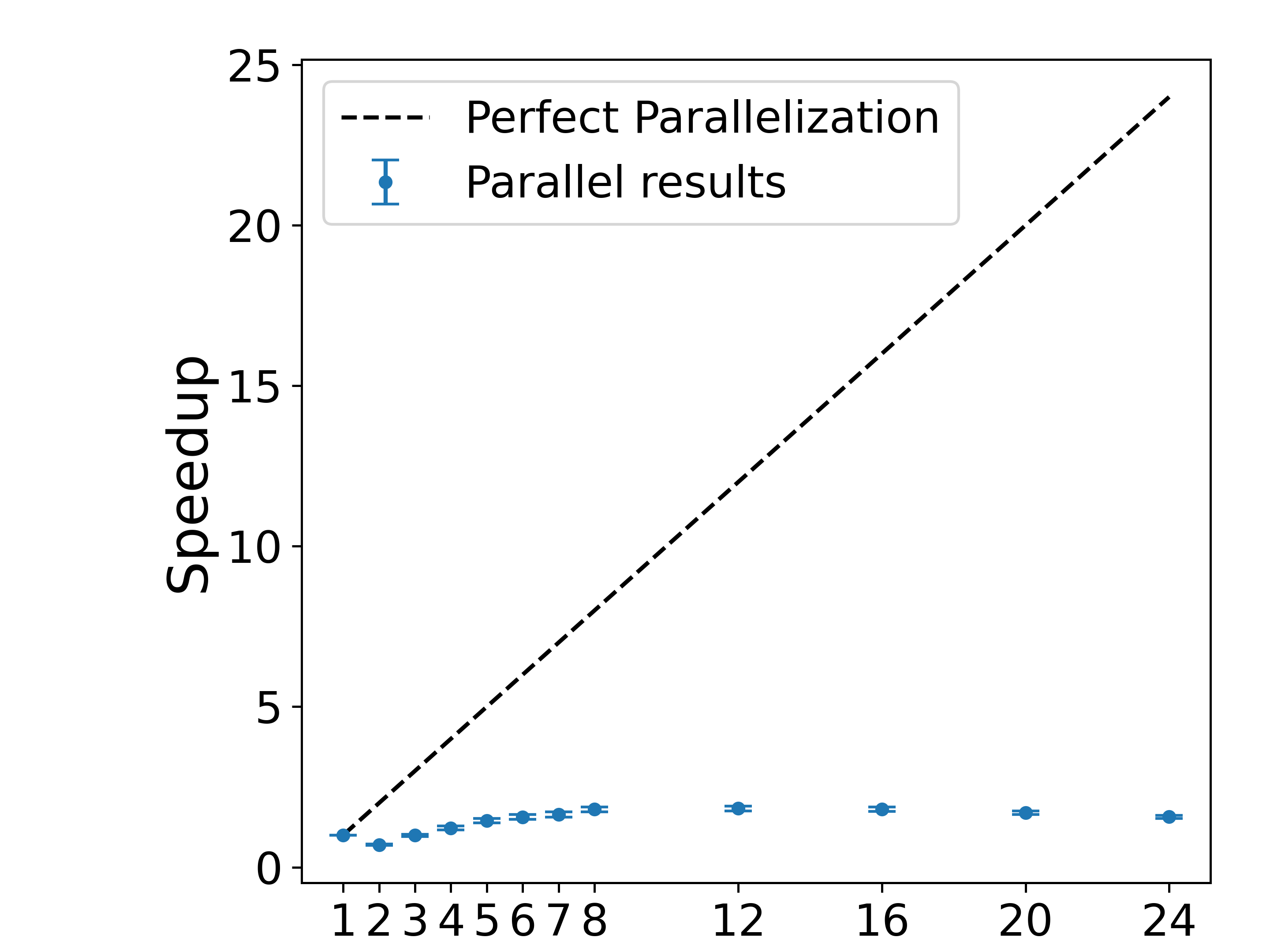}\caption{\mynote{Speedup curve for parallelization for processing liquid crystal images. Perfect speedup line shown for reference.}}\label{fig:Speedup_LC}
\end{center}
\end{figure}

Also, the potential limitations of parallelization are noticed when considering the microscopy case. For larger random fields for benchmarking, more cores yielded significantly faster performance, whereas in the liquid crystal case study, more cores did not yield favorable speedup. This is seen with a peak efficiency at about 12 cores before dropping as more cores were used (Figure \ref{fig:Speedup_LC}). This analysis does not consider simply using Algorithm \ref{alg:2D_contributions_serial} for each individual image, but running the dispatch of the image analysis in parallel. This highlights the importance to design a computational framework that accurately addresses data analysis and data structure from case to case. This is especially relevant in the case of designing lightweight, standalone sensors with limited computational capabilities.

The molecular dynamics study shows similar trends as the microscopy study, indicating the data was small enough that parallelization did not realize the same benefit as random 3D fields of larger size. However, this framework is capable a handling molecular dynamics simulations which may include thousands or millions of molecules. The granularity of the grid (20x20x20) could be expanded to have more fine-grained analysis on larger systems and see significant speed-up similar to that found in the random 3D fields analyzed earlier. Also, any 3D field data that can be translated to a cubical lattice may be analyzed in the same way. For instance, computational fluid dynamics data with millions of finite elements could be analyzed for trends in EC which could transform large-scale, fine-grained data into a feasible size for input into a prediction algorithm without losing important topological information. For reference, a couple of computational fluid dynamics simulations were analyzed to demonstrate scalability: first, a 302x302x302 element simulation of heptane gas undergoing combustion (The University of Utah Center for the Simulation of Accidental Fires and Explosions), took 405 seconds for {\tt GUDHI}, and 3.5 seconds for Algorithm \ref{alg:3D_contributions_serial} in parallel with 4 cores; and second, simulation of duct flow \cite{Atzori_CFD} of size 193x194x1000 took GUDHI 673 seconds, whereas Algorithm \ref{alg:3D_contributions_serial} in parallel with 4 cores took 5.5 seconds. Clearly, scaling advantages exist in these data-rich applications.

For the hyperspectral image study, the size of the images still indicated that parallelizing the identification of vertex contributions was economical. The average size of images in this data set was 6.88 MV, meaning it takes about 100 seconds to process a single hyperspectral image with GUDHI, 3 seconds in serial with Algorithm \ref{alg:2D_contributions_serial} in 3D mode, and only about 0.2 seconds when using \ref{alg:2D_contributions_parallel} in 3D mode with 24 cores. Applications where sensors are developed with a hyperspectral camera may require processing speeds the allow data analysis at speeds faster than those presented in this work. However, in certain applications, such as monitoring pharmaceutical powder composition, standard sampling times for single-sample analysis (only spectral dimension; no 2D image component) near infrared sensors can be on the order of 5 - 15 seconds \cite{NIR_sensor}. On the other hand, commercially available hyperspectral cameras can achieve frame rates that could facilitate real-time data \cite{HSI_specim}. Once again, understanding the needs of the system and the dynamic response to control actions should be considered case-to-case when implementing topology-informed control schemes to hyperspectral data.

An important distinction with respect to hyperspectral imaging that we used the raw data for the case study. Performing dimensionality reduction in hyperspectral image is not uncommon, for instance, using principal component analysis to reduce the number of wavelengths used in the spectral dimension \cite{HSI_review}. Reducing the number of wavelengths through dimensionality reduction or selecting regions that are known to contain the most important topological information could reduce computation time enough to process these 3D hyperspectral images in real time as well.

Also, the output of this tool is not just the EC curve, it is the vertex contribution map over the filtration values. Thus, by simply changing the weights each vertex contributes to an overall topological descriptor, the same vertex contribution map can be used to compute any number of topological descriptors as a function of vertex contributions. Also, connectivity need not be assumed in the beginning, as the connectivity only impacts the weights, not the vertex types. This allows for analyses to identify that alternate definitions of connectivity may be more suitable for a given physical system.

With this being said, sensitivity of the system to connectivity can be explored alongside sensitivity of topological descriptors to image resolution, number of filtration values, or image preprocessing techniques. It is commonplace to blur images or treat images with another convolutional operator while utilizing neural networks in machine learning. However, in this work, we use the raw data for both applications. Since the tool is scalable and allows for large-scale analysis, performing studies to understand which combination of resolution, number of filtrations, preprocessing technique, and connectivity type may lead to more topologically-based physical intuition of chemical systems discerned through field/image analysis.

\section{Conclusions}

In this work, we presented algorithms addressing the scalable computation of vertex contributions in field data. Ultimately, the tool was tested using EC, but the analysis that can be done with vertex contribution maps is not limited just to EC. The speedup was significant, 2 to 3 orders of magnitude, when compared with a common package used for topological persistence in Python ({\tt GUDHI}). Also, recent advancement of the {\tt CHUNKYEuler} software show similar timing capabilities, indicating that methods that address changes in topological contributions are much more scalable than counterparts that compute topological characteristics at each level of a filtration set.

Future research directions could address the usage of vertex contribution maps for more complex topological descriptors, such as the fractal dimension. Also, generalizing these methods to larger dimensions (4D data, such as hyperspectral imaging with a temporal dimension) requires either algorithms to generate or research into vertex contribution maps in higher dimensions. In terms of low memory methods, {\tt CHUNKYEuler} uses chunks of each image, whereas the low memory version of algorithms \ref{alg:2D_contributions_serial} and \ref{alg:2D_contributions_serial} reads only the minimum data required to compute the EC curve. Exploration into the optimal decomposition scheme for a cubical field/image could be an interesting direction for research, as having minimum data representations is good for memory scaling, but results in a slow down of just over an order of magnitude.

Also, generalizing vertex contributions to regular, non-cubical lattices, or even more generally to Voronoi cells, with sets of vertex contributions corresponding to the degree of that vertex could be impactful for data that is not the classical cubical layout that physical fields/images possess. Ultimately, if an efficient method exists to compute vertex contribution maps of these more abstract structures, one could calculate a plethora of topological and general system descriptors to control or characterize complex and more abstract physical systems in real time.

\section{Acknowledgements}
This research was supported by the U.S. National Science Foundation through the University of Wisconsin Materials Research Science and Engineering Center (DMR-2309000) and via grant IIS-1837812. 


\newpage

\appendix

\section{3D Algorithms}
\begin{algorithm}
\caption{Bitmap contribution (3D, serial). Parallel mode runs lines \ref{line:3D_parallel_start} to \ref{line:3D_parallel_stop} on different threads.} \label{alg:3D_contributions_serial}
\begin{algorithmic}[1]
	\STATE Field $\gets$ read(data file)
	\STATE$w \gets$ field width
	\STATE$h \gets$ field height
	\STATE$d \gets$ field depth
	\STATE$M \gets$ largest filtration value
	\STATE$q_{k,\text{in}}, q_{k,\text{out}} \gets$ Zeros of length M+2 $\forall k \in \{0, 1, ..., 42\}$
	\FORALL{$i < h + 1$} \label{line:3D_parallel_start}
		\FORALL{$j < w + 1$}
			\FORALL{$k < d + 1$}
				\STATE data $\gets [M+1, M+1, M+1, M+1, M+1, M+1, M+1, M+1]$ 
				\STATE data $\gets$ \text{fill\_data\_3D} $\left(i, j, k\right)$ (Algorithm \ref{alg:fill_data_3D})
				\STATE pos $\gets$ argsort(data)
				\STATE adjacency $\gets$ compute $\left[f_\text{adj}\left(\text{pos}[0, 1]\right), f_\text{adj}\left(\text{pos}[0, 1, 2]\right), f_\text{adj}\left(\text{pos}[0, 1, 2, 3]\right)\right]$
				\STATE adjacency$_\text{empty}$ $\gets$ compute $\left[f_\text{adj}\left(\text{pos}[7, 6]\right), f_\text{adj}\left(\text{pos}[7, 6, 5]\right)\right]$
				\STATE $q_{10,\text{in}}[\text{data}[0]]\texttt{++}, q_{10,\text{out}}[\text{data}[1]]\texttt{++}$
				\IF{adjacency[0] $< \sqrt{2}$}
					\STATE $q_{20,\text{in}}[\text{data}[1]]\texttt{++}, q_{20,\text{out}}[\text{data}[2]]\texttt{++}$
				\ELSIF{adjacency[0] $< \sqrt{3}$}
					\STATE $q_{21,\text{in}}[\text{data}[1]]\texttt{++}, q_{21,\text{out}}[\text{data}[2]]\texttt{++}$
				\ELSE
					\STATE $q_{22,\text{in}}[\text{data}[1]]\texttt{++}, q_{22,\text{out}}[\text{data}[2]]\texttt{++}$
				\ENDIF
				\IF{adjacency[0] $< 1 + \sqrt{2} + \sqrt{3}$}
					\STATE $q_{30,\text{in}}[\text{data}[2]]\texttt{++}, q_{30,\text{out}}[\text{data}[3]]\texttt{++}$
				\ELSIF{adjacency[0] $< \sqrt{3}$}
					\STATE $q_{31,\text{in}}[\text{data}[2]]\texttt{++}, q_{31,\text{out}}[\text{data}[3]]\texttt{++}$
				\ELSE
					\STATE $q_{32,\text{in}}[\text{data}[2]]\texttt{++}, q_{32,\text{out}}[\text{data}[3]]\texttt{++}$
				\ENDIF
				\STATE $\vdots$
				\STATE Continue similarly for $q_{40}$ through $q_{80}$. Use adjacency thresholds in Table \ref{tab:3D_bitmap_contributions}
				\STATE $\vdots$
			\ENDFOR
		\ENDFOR
	\ENDFOR \label{line:3D_parallel_stop}
\end{algorithmic}
\end{algorithm}

\begin{algorithm}
\caption{Filling data for 3D field.} \label{alg:fill_data_3D}
\begin{algorithmic}[1]
	\STATE $(i, j, k) \gets$ input values
	\STATE data $\gets$ input data
				\IF{$i > 0 \land j > 0 \land k > 0$}
					\STATE data$[0] \gets$ Field$(i - 1, j - 1, k-1)$
				\ENDIF
				\IF{$i > 0 \land j > 0 \land k < d$}
					\STATE data$[4] \gets$ Field$(i - 1, j - 1, k)$
				\ENDIF
				\IF{$i > 0 \land j < w \land k > 0$}
					\STATE data$[2] \gets$ Field$(i - 1, j, k-1)$
				\ENDIF
				\IF{$i > 0 \land j < w \land k < d$}
					\STATE data$[6] \gets$ Field$(i - 1, j, k)$
				\ENDIF
				\IF{$i < h \land j > 0 \land k > 0$}
					\STATE data$[1] \gets$ Field$(i, j - 1, k-1)$
				\ENDIF
				\IF{$i < h \land j > 0 \land k < d$}
					\STATE data$[5] \gets$ Field$(i, j - 1, k)$
				\ENDIF
				\IF{$i < h \land j < w \land k > 0$}
					\STATE data$[3] \gets$ Field$(i, j, k-1)$
				\ENDIF
				\IF{$i < h \land j < w \land k < d$}
					\STATE data$[7] \gets$ Field$(i, j, k)$
				\ENDIF
\end{algorithmic}
\end{algorithm}

\end{document}